\documentclass[draft]{agujournal2019}
\usepackage[T1]{fontenc}  
\usepackage{url} 
\usepackage{soul}
\draftfalse

\usepackage{booktabs}       
\usepackage{tabularx}
\usepackage{arydshln}
\usepackage{multicol}
\usepackage{multirow}
\usepackage{amsfonts}       
\usepackage{nicefrac}       
\usepackage{microtype}      
\usepackage{graphicx}
\usepackage{subcaption}     
\usepackage{amsmath}
\usepackage{siunitx}        
\usepackage{pifont}         
\usepackage{xcolor}
\usepackage{comment}

\definecolor{carnelian}{rgb}{0.7, 0.11, 0.11}
\definecolor{cadmiumgreen}{rgb}{0.0, 0.42, 0.24}
\definecolor{cnxtblue}{RGB}{137, 165, 214}
\definecolor{poolgreen}{RGB}{0, 104, 55}
\definecolor{convtbrown}{RGB}{140, 98, 57}
\definecolor{convred}{RGB}{193, 39, 45}
\definecolor{grupurp}{RGB}{194, 161, 204}
\definecolor{convblue}{RGB}{0, 0, 255}

\newcolumntype{P}[1]{>{\centering\arraybackslash}p{#1}}
\newcolumntype{R}{>{\raggedleft\arraybackslash}X}
\newcolumntype{C}{>{\centering\arraybackslash}X}

\newcommand\Tstrut{\rule{0pt}{2.6ex}}         
\newcommand\Bstrut{\rule[-0.9ex]{0pt}{0pt}}   

\newcommand{\revA}[1]{{\color{black}#1}}
\newcommand{\revB}[1]{{\color{black}#1}}

\usepackage{etoolbox}
\makeatletter
\patchcmd{\hyper@makecurrent}{%
    \ifx\Hy@param\Hy@chapterstring
        \let\Hy@param\Hy@chapapp
    \fi
}{%
    \iftoggle{inappendix}{
        \@checkappendixparam{chapter}%
        \@checkappendixparam{section}%
        \@checkappendixparam{subsection}%
        \@checkappendixparam{subsubsection}%
        \@checkappendixparam{paragraph}%
        \@checkappendixparam{subparagraph}%
    }{}%
}{}{\errmessage{failed to patch}}
\newcommand*{\@checkappendixparam}[1]{%
    \def\@checkappendixparamtmp{#1}%
    \ifx\Hy@param\@checkappendixparamtmp
        \let\Hy@param\Hy@appendixstring
    \fi
}
\makeatletter
\newtoggle{inappendix}
\togglefalse{inappendix}
\apptocmd{\appendix}{\toggletrue{inappendix}}{}{\errmessage{failed to patch}}

\journalname{Journal of Advances in Modeling Earth Systems (JAMES)}

\begin{document}

%
%


\title{Advancing Parsimonious Deep Learning Weather Prediction using the HEALPix Mesh}

%
%




\authors{Matthias Karlbauer\affil{1}, Nathaniel Cresswell-Clay\affil{2}, Dale R. Durran\affil{2},\\ Raul A. Moreno\affil{2}, Thorsten Kurth\affil{3}, Boris Bonev\affil{3}, Noah Brenowitz\affil{4}, and Martin V. Butz\affil{1}}


\affiliation{1}{Neuro-Cognitive Modeling Group, Department of Computer Science, University of Tübingen, Tübingen, Germany}
\affiliation{2}{Department of Atmospheric Sciences, University of Washington, Seattle, WA, USA}
\affiliation{3}{NVIDIA Switzerland AG, Zürich, Switzerland}
\affiliation{4}{NVIDIA Corporation, Seattle, USA}




\correspondingauthor{Dale R. Durran}{drdee@uw.edu}



\begin{keypoints}
\item \revB{The model forecasts 7 atmospheric variables, an order of magnitude less than that used in state-of-the-art ML weather forecast models.}
\item Forecasts are generated on the HEALPix mesh, facilitating the development of location invariant convolution kernels.
\item Without converging to climatology, the model produces realistic \revB{atmospheric states in 365-day iterative rollouts}.
\end{keypoints}

%
%

%
%


\begin{abstract}
We present a parsimonious deep learning weather prediction model \revB{to forecast seven atmospheric variables with 3-h time resolution for up to one-year lead times on a 110-km global mesh using the Hierarchical Equal Area isoLatitude Pixelization (HEALPix)}.
In comparison to state-of-the-art \revB{(SOTA)} machine learning \revB{(ML}) weather forecast models, such as Pangu-Weather and GraphCast, our DLWP-HPX model uses coarser resolution and far fewer prognostic variables.
Yet, at one-week lead times, its skill is only about one day behind \revB{both SOTA ML forecast models and the SOTA} numerical weather prediction model from the European Centre for Medium-Range Weather Forecasts. 
We report \revB{several improvements in} model design\revB{, including}  switching from the cubed sphere to the HEALPix mesh, inverting the channel depth of the U-Net, and introducing gated recurrent units (GRU) on each level of the U-Net hierarchy.
The consistent east-west orientation of all cells on the HEALPix mesh facilitates the development of location-invariant convolution kernels that successfully propagate weather patterns across \revB{the globe without requiring separate kernels for the polar and equatorial faces of the cube sphere.} 
Without any loss of spectral power after \revB{the first} two days, the model can be unrolled autoregressively for hundreds of steps into the future to generate realistic states of the atmosphere that respect seasonal trends, as showcased in one-year simulations.
\end{abstract}

\section*{Plain Language Summary}

Weather forecasting traditionally relies on numerical weather prediction models that solve physical equations to simulate the evolution of the atmosphere.
Such numerical models are compute intensive, and their performance is increasingly challenged by less compute demanding but still highly sophisticated machine learning (ML) approaches.
Yet, a downside for many of these new ML models is \revB{they tend to drift away from climatology while producing excessively smoothed fields if they are iteratively stepped forward for several months.}
Here, a parsimonious machine learning model is developed to forecast just \revB{7 atmospheric variables that can be stepped forward to give realistic weather patterns over a full year. 
Despite using at least a factor of 10 less variables than the 67 to 227 in the best ML models, our model generates eight-day forecasts with errors that are only a day behind those from state-of-the-art ML forecasts. Our model provides a path toward sub-seasonal and seasonal forecasting that could potentially improve planning for agriculture, water resources, disaster preparedness, and energy production} 

%
%

\section{Introduction}\label{sec:introduction}

Four years ago, \citeA{weyn2019can} posed the question ``Can machines learn to predict the weather?'' and demonstrated that 
data driven convolutional neural networks can forecast the evolution of the \SI{500}{hPa} surface much better than the alternative dynamical model, the barotropic vorticity equation, which was used in the first numerical weather prediction (NWP) model \cite{Charney1950}. An extremely rapid evolution of deep learning weather prediction (DLWP) models followed, culminating in the recent Pangu-Weather \cite{bi2023pangu} and GraphCast models \cite{lam2022graphcast}, which outperform the \revA{deterministic forecast from} the state-of-the-art Integrated Forecast System (IFS) of the European Centre for Medium-Range Weather Forecasts (ECMWF).

NWP has continuously improved over the seven decades since the first barotropic model forecast \cite{benjamin2019100}. Current state-of-the-art models typically provide skillful predictions of global weather patterns at effective grid point spacings of roughly $\SI{0.1}{^\circ}$ of latitude (about \SI{10}{km}) through at least seven days of forecast lead time \cite{bauer2015quiet}. The computational effort required to generate such global high-resolution forecasts is enormous and only available at a handful of advanced dedicated centers. 
Ensemble forecasts, which provide an important way to account for uncertainty by generating a set of equally plausible predictions and extend the limit of skillful forecasts beyond that of a single deterministic model run, are also limited by the computational burden of high-resolution NWP to about 50 members \cite{palmer2019ecmwf}.

Global NWP models represent 3D fields as sets of nested spherical shells in which the distance between each shell is the local vertical grid spacing.  On every time step, the ECMWF Integrated Forecasting System (IFS), as configured for sub-seasonal forecasting, updates \revA{10} prognostic 3D variables defined at 91 vertical levels. Along with surface pressure, this totals to \revA{over 900} spherical shells of data. Here, we use ``spherical shell of data'' to describe a single variable defined at a single vertical level on a spherical shell covering the globe. The large number of spherical shells of data (combined with the fine horizontal resolution) in NWP models is required to produce acceptably accurate numerical solutions to the equations governing atmospheric motions. The data at each individual point, however, cannot be independently perturbed while maintaining a meteorologically relevant atmospheric state. For example, on horizontal scales larger than about \SI{10}{km}, the temperatures throughout a vertical column and the heights of constant pressure surfaces must satisfy hydrostatic balance. 

The actual number of independent degrees of freedom required to represent the predictable components of the global atmosphere is unknown, but it clearly decreases with increasing forecast lead times \cite{lorenz69}. GraphCast \cite{lam2022graphcast}, for example, has achieved success at lead times as short as \SI{6}{h} with 227 spherical shells of data. It can produce forecasts using much less computation time than the ECMWF IFS, but it still requires large computing resources for training: 3 weeks using 32 TPU 4 processors. Pangu-Weather \cite{bi2023pangu} cuts the number of spherical shells by almost 2/3 to 69. \revA{The spherical Fourier neural operator (SFNO) version of FourCastNet compared with the IFS in \citeA{bonev2023sfno} uses 73 spherical shells of data.} Here, we take this reduction much farther, presenting a parsimonious DLWP model that uses just 7 spherical shells of data to efficiently provide forecasts approaching the skill of ECMWF. While not as accurate as GraphCast or Pangu-Weather \revA{for medium range forecasts with lead times less than two weeks, we demonstrate that our model generates far less bias in forecasts of 500-hPa height in one-year iterative forecasts. In addition, our model is potentially better suited for research applications such as computing the sensitivities of its compact state vector to custom diagnostic functions by backpropagation.} 

In contrast to many of the recent DLWP architectures, our approach relies on convolutional neural networks (CNN), building on early work by \citeA{scher2018predicting} and \citeA{weyn2019can} and the U-Net configuration in \citeA{weyn2020improving} and \citeA{weyn2021sub}. Here, we document substantial improvements over \citeA{weyn2021sub}, obtained
by replacing the cubed sphere data representation with the HEALPix mesh, which is widely employed in astronomy \cite{gorski2005healpix}. In addition, we improve the former model by implementing physically motivated modifications in form of residual connections, recurrent modules, and inverting the channel depth as compared with a standard U-Net.

\section{Related Work}\label{sec:related_work}

Pioneering efforts to create machine learning models to forecast the weather from reanalysis or general circulation model (GCM) output include the dense neural network of \citeA{Dueben2018design} and the CNN models of \citeA{Scher2019nn_GCM} and \citeA{weyn2019can}, all of which employed latitude longitude (lat-lon) meshes. \citeA{weyn2020improving} obtained significantly improved forecasts by switching to a cubed sphere mesh with a CNN in the standard U-Net architecture \cite{ronneberger2015u}. Their model was capable of generating realistic weather patterns when stepped forward for a full year (730 \SI{12}{h} steps). Retaining the cubed sphere, \citeA{weyn2021sub} produced forecasts out to sub-seasonal time scales using large multi-model ensembles, and \citeA{lopez2022global} migrated from the U-Net into a U-Net 3+ architecture \cite{huang2020unet}\revA{---which adds connections between multiple hierarchical levels in the U-Net---}to generate forecasts of extreme surface temperatures.

Returning to the lat-lon mesh, \citeA{rasp2021data} demonstrated that a deep Resnet could be pre-trained on GCM data and then fine-tuned by transfer learning on ERA5 data to produce up to 5-day forecasts at coarse $\SI{5.65}{^\circ}$ grid spacing. Building on transformer models from computer vision \cite{dosovitskiy2020image,guibas2021efficient}, 
\citeA{pathak2022fourcastnet} and \citeA{kurth2022fourcastnet} used Fourier neural operators \cite{li2020fourier} to develop FourCastNet on a $\SI{0.25}{^\circ}$ lat-lon mesh to generate forecasts approaching the accuracy of ECMWF's IFS. FourCastNet was not, however, capable of stable long-lead-time autoregressive rollouts. This difficulty was overcome by switching from 2D Fourier modes on a lat-lon mesh to spherical harmonic functions \citeA{bonev2023sfno}. The resulting SFNO model eliminated much of the vision transformer architecture while improving accuracy and remaining stable for one-year forecasts.  

Again on a 5.65$^{\circ}$ lat-lon mesh, \citeA{hu2022swinvrnn} used a shifted window (Swin) transformer \cite{liu2021swin} to produce single forecasts as well as ensembles generated by perturbing the latent state using samples from their learned distribution. \citeA{bi2023pangu} also applied Swin transformers on a lat-lon mesh, but used a fine $\SI{0.25}{^\circ}$ lat-lon grid spacing, 3D transformers,  and included latitude and longitude fields as input to train a ``3D Earth-specific transformer'' at four different forecast lead times of \SI{1}{}, \SI{3}{}, \SI{6}{}, and \SI{24}{h}, which are used in combination to span an arbitrary hourly forecast period with minimal model steps. If the ECMWF IFS NWP forecasts are averaged to the coarser $\SI{0.25}{^\circ}$ lat-lon mesh, Pangu-Weather outperforms NWP on several metrics.

In contrast to the preceding approaches, graph neural networks \cite{gori2005new,scarselli2008graph,kipf2016semi,battaglia2018relational,pfaff2020learning} where applied on icosahedral meshes at course resolution by \citeA{keisler2022forecasting}
and at fine resolution in the GraphCast model \cite{lam2022graphcast}. Similarly to Pangu-Weather, GraphCast appears to outperform the coarsened ECMWF IFS forecast on several metrics.

\section{Methods}\label{sec:methods}

\subsection{Data}\label{sec:methods.data}

\subsubsection{Choice of Variables}
Beginning with the same six prognostic variables used in \citeA{weyn2021sub}---geopotential height at \SI{1000}{hPa} and \SI{500}{hPa} ($Z_{1000}$, $Z_{500}$),\footnote{The related variable in the ERA5 dataset is geopotential and named $z$, whereas the geopotential height, typically referred to as $Z$, represents the actual height above sea level of the respective pressure surface and is obtained by dividing geopotential by the gravitational constant.} \SI{700}{hPa} to \SI{300}{hPa} thickness ($\tau_{700-300}$) defined as $Z_{300}-Z_{700}$, temperature at \SI{2}{m} height above ground ($T_{2m}$), temperature at \SI{850}{hPa} ($T_{850}$), and total column water vapor ($TCWV$)---we add $Z_{250}$ based on its importance in the model of \citeA{rasp2021data} and to provide an upper tropospheric variable. As in \citeA{weyn2021sub}, three prescribed fields are also provided: topographic height, land-sea mask, and top-of-atmosphere (TOA) incident solar radiation. \revA{We do not include prescribed or predicted sea-surface temperature or surface fluxes above the land or ocean.} {\it No} specific information about position on the globe, such as latitude and longitude, is provided. Three-hourly data from the ERA5 reanalysis \cite{hersbach2020era5} provide training data from 1979-2012, a validation set from 2013-2016, and a test set from 2017-2018.

\subsubsection{HEALPix Mesh}
\begin{figure}
    \centering
    \includegraphics[width=\textwidth]{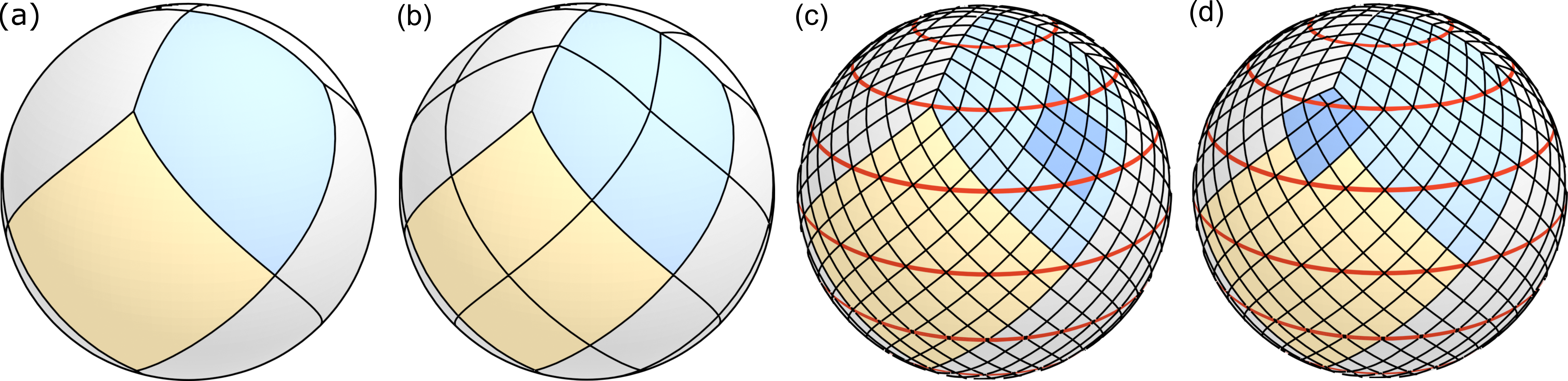}
    \caption{Division of the sphere into twelve faces according to the HEALPix. Four faces to represent either the northern (blue) and southern extratropics, while four more faces arrange around the equator to represent the tropics (yellow). Each face can be subdivided into patches with divisions along the side of each face given by powers of two. The sphere in (a) has a pixel-count of one per face side; we call it \texttt{hpx1}. The sphere in (b) counts two pixels per side (\texttt{hpx2}), whereas the two spheres in (c) and (d) have eight pixels per side, i.e., \texttt{hpx8}. Several latitude lines in red emphasize the iso-latitudinal arrangement of the patches. The saturated blue area depicts a $3\times3$ stencil, as applied by a standard convolution. To apply the $3\times3$ stencil at the top corner of the equatorial faces, i.e., stencil position in (d), we fill in the missing corner patch with the average of the values in the two adjacent patches on the extratropical faces.}
    \label{fig:healpix}
\end{figure}
We discretize all fields using the Hierarchical Equal Area isoLatitude Pixelization (HEALPix) \cite{gorski2005healpix}. As
depicted in \autoref{fig:healpix}, a HEALPix mesh is formed by dividing the sphere into twelve equal-area diamond-shaped faces, with four faces lying in the northern and southern hemispheres, and four in the tropics.
According to \citeA{gorski2005healpix}, the HEALPix mesh has three important properties. \emph{(1) Hierarchical structure of the database:} Each of the twelve base faces can be progressively subdivided into smaller patches. \emph{(2) Equal areas for the discrete elements of the partition:} All patches are the same size. \emph{(3) Isolatitude distribution for the discrete area elements on the sphere:} The patches line up with lines of latitude, facilitating the computation of zonal averages and one-dimensional zonal spectra. Importantly, this last property makes the HEALPix mesh an ``east is to the right'' grid, which facilitates the training of \revB{a single set of position invariant convolutional} kernels to capture the motion of typical weather disturbances, as discussed in \autoref{sec:results.2week}.

The HEALPix can be considered a graph and does not allow a seamless application of convolution operations. Thus, \citeA{perraudin2019deepsphere} explicitly define a graph from the HEALPix---by connecting adjacent neighbors with weighted edges---and perform a graph convolution to classify weak lensing maps from cosmology. In a different approach, \citeA{krachmalnicoff2019convolutional} classify digits and determine cosmic parameters from simulated cosmic microwave background maps. They apply 1D convolutions to the flattened HEALPix data with a kernel size $k$ and stride $s$ both equal to 9, appending a zero to those cases where only seven instead of eight neighbors are defined (top corner of the tropical faces). In contrast, we treat the twelve faces as distinct images and pad their boundaries using data from neighboring faces to allow the computation of 2D convolutions and averaging operators directly, as detailed in   \autoref{app:sec:deep_learnin_on_the_healpix}. To accelerate the padding operation, we have implemented a custom CUDA kernel, which is available in our repository.\footnote{\url{https://github.com/CognitiveModeling/dlwp-hpx}}

The grid spacing, or shortest inter-node spacing, on the HEALPix mesh is the diagonal distance between a pair of nodes on adjacent latitude lines. Denoting a HEALPix mesh with $n$ divisions along one side of the original 12 faces as HPX$n$. The grid spacing is approximately 220 km ($\approx 2{^\circ}$) for HPX32 and 
110 km ($\approx 1{^\circ}$) for HPX64.\footnote{We provide download explanations and projection scripts in our repository. The 3D HEALPix figures are drawn in Blender 3.4.1; respective Blender files are provided in the repository too.}

\subsection{Machine Learning Architecture}\label{sec:methods.model}

\begin{figure}
    \centering
    \includegraphics[width=\textwidth]{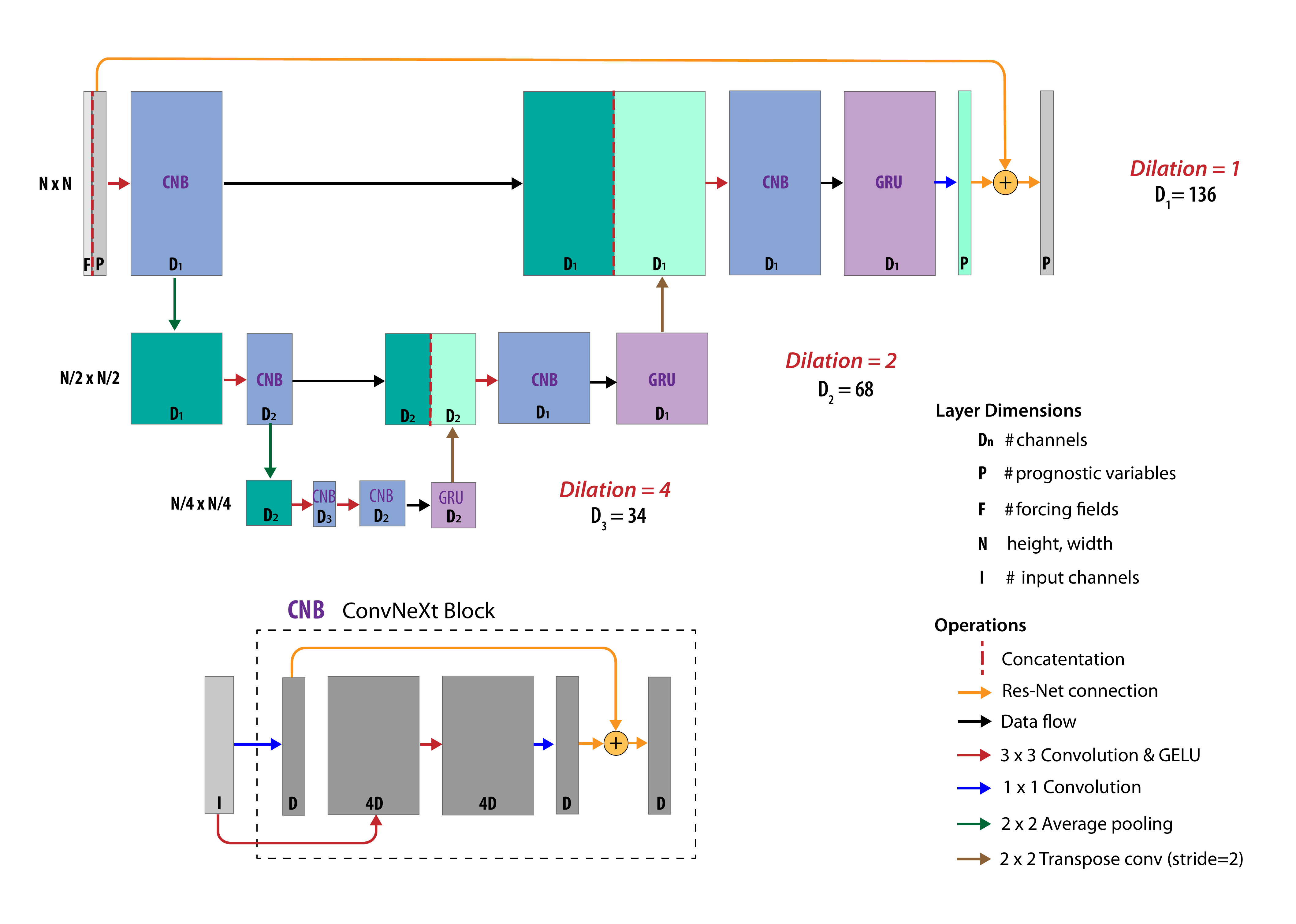}
    \caption{\revB{Schematic representation of our DLWP-HPX architecture as a sequence of operations on layers (see legend). Individual layers are labeled by their channel depth, with $D_1=136$, $D_2=68$, and $D_3=34$ being associated with the first convolutions in each of the three U-Net levels. Each ConvNeXt block (blue) is replaced by the layers and operations shown in the inset labeled CNB, with generic depths $D$ and $I$ determined by the channel depth of the input and the labeled value of $D_n$. The purple blocks labeled GRU denote convolutional Gated Recurrent Unit layers, which are augmented with $1\times 1$ spatial convolutions. Other layers evaluated by the encoder are shown as dark green, while those evaluated by the decoder are shown as light green.}}
    \label{fig:model}
\end{figure}

Relating to Tobler's first law of geography: ``All things are related, but nearby things are more related than distant things.'' \cite{tobler1970computer}, we mostly retain the comparably simple U-Net structure from \citeA{weyn2020improving}.
U-Nets \cite{ronneberger2015u} are hierarchically structured feed-forward convolutional neural networks that were originally proposed for segmenting biomedical images. 
The U-Net structure proposed here introduces several physically motivated advancements to the vanilla U-Net used by \citeA{weyn2021sub} for sub-seasonal forecasting. 
\revB{Our final model configuration is visualized as a sequence of operations on layers or a block of layer operations in \autoref{fig:model}.
The latter case is indicated by CNB or GRU, which refer to \texttt{ConvNeXt}- and \texttt{GRU}-blocks (cf. \autoref{sec:methods.model.residual_prediction} and \autoref{sec:methods.model.recurrent_modules} for explanations), respectively.
Details of the \texttt{ConvNeXt}-block structure are also visualized.
\texttt{GRU}-blocks augment the respective layer with a recurrent processing mechanism (cf., \autoref{sec:methods.model.recurrent_modules}). 
\autoref{app:tab:layer_parameters} specifies the respective parameter settings. 
Color codes for the operations in \autoref{app:tab:layer_parameters} approximate those used in the model schematic in \autoref{fig:model}. For example, the operations in red are $3\times 3$ convolutions followed by GELU activation functions. Residual connections are only reported in \autoref{app:tab:layer_parameters} 
if they contribute to the parameter count when implementing a $1\times1$ convolution to adjust the channel depth.}
In the following, we describe the incremental advancements that we add to our model.
\begin{table}
    \centering
    \caption{\revB{CNN architecture as a sequence of operations on layers; $c_{\rm in}$, $k$, $s$ and $d$ denote the number of input channels, kernel size, stride, and dilation. Output shape is face$\times$height$\times$width$\times$channels. Dashed line separates model's encoder (above) and decoder (below). ``Concat" implements skip connections by appending the state in parenthesis, numbered earlier, to the output of the previous layer. The result from the orange 1$\times$1 convolution at the beginning of most ConvNeXt blocks is added to the corresponding output channel to form a residual connection.}}
    \label{app:tab:layer_parameters}
    \footnotesize
    \begin{tabularx}{\textwidth}{lCCCCccrrr}
        \toprule
         & & & & & Receptive & & \multicolumn{3}{c}{Parameter count} \\
        \cmidrule{8-10}
        Layer & $c_{\operatorname{in}}$ & $k$ & $s$ & $d$ & Field & Output shape & Weights & Biases & Total\\
        \midrule
        
        \multicolumn{4}{l}{\texttt{\color{cnxtblue}ConvNeXt}} & & & & & & \\
        \hspace{0.3cm}\texttt{\color{orange}Conv2d} & \phantom{0}18 & 1 & 1 & 1 & $1\times1$ & (12, 64, 64, 136) & 2\,448 & 136 & 2\,584 \\
        \hspace{0.3cm}\texttt{\color{convred}Conv2d} & \phantom{0}18 & 3 & 1 & 1 & $3\times3$ & (12, 64, 64, 544) & 88\,128 & 544 & 88\,672 \\
        \hspace{0.3cm}\texttt{\color{convred}Conv2d} & 544 & 3 & 1 & 1 & $5\times5$ & (12, 64, 64, 544) & 2\,663\,424 & 544 & 2\,663\,968 \\
        \hspace{0.3cm}\texttt{\color{convblue}Conv2d} (1) & 544 & 1 & 1 & 1 & $5\times5$ & (12, 64, 64, 136) & 73\,984 & 136 & 74\,120 \\
        
        \Tstrut
        \texttt{\color{poolgreen}AvgPool2d} & 136 & 2 & 2 & --- & $6\times6$ & (12, 32, 32, 136) & 0 & 0 & 0 \\
        \texttt{\color{cnxtblue}ConvNeXt} & & & & & & & & & \\
        \hspace{0.3cm}\texttt{\color{orange}Conv2d} & 136 & 1 & 1 & 1 & $6\times6$ & (12, 32, 32, \phantom{0}68) & 9\,248 & 68 & 9\,316 \\
        \hspace{0.3cm}\texttt{\color{convred}Conv2d} & 136 & 3 & 1 & 2 & $14\times14$ & (12, 32, 32, 272) & 332\,928 & 272 & 333\,200 \\
        \hspace{0.3cm}\texttt{\color{convred}Conv2d} & 272 & 3 & 1 & 2 & $22\times22$ & (12, 32, 32, 272) & 665\,856 & 272 & 666\,128 \\
        \hspace{0.3cm}\texttt{\color{convblue}Conv2d} (2) & 272 & 1 & 1 & 1 & $22\times22$ & (12, 32, 32, \phantom{0}68) & 18\,496 & 68 & 18\,564 \\
        
        \Tstrut
        \texttt{\color{poolgreen}AvgPool2d} & \phantom{0}68 & 2 & 2 & --- & $24\times24$ & (12, 16, 16, \phantom{0}68) & 0 & 0 & 0 \\
        \texttt{\color{cnxtblue}ConvNeXt} & & & & & & & & & \\
        \hspace{0.3cm}\texttt{\color{orange}Conv2d} & \phantom{0}68 & 1 & 1 & 1 & $24\times24$ & (12, 16, 16, \phantom{0}34) & 2\,312 & 34 & 2\,346 \\
        \hspace{0.3cm}\texttt{\color{convred}Conv2d} & \phantom{0}68 & 3 & 1 & 4 & $56\times56$ & (12, 16, 16, 136) & 83\,232 & 136 & 83\,368 \\
        \hspace{0.3cm}\texttt{\color{convred}Conv2d} & 136 & 3 & 1 & 4 & $88\times88$ & (12, 16, 16, 136) & 166\,464 & 136 & 166\,600 \\
        \hspace{0.3cm}\texttt{\color{convblue}Conv2d} & 136 & 1 & 1 & 1 & $88\times88$ & (12, 16, 16, \phantom{0}34) & 4\,624 & 34 & 4\,658 \Bstrut\\
        
        \hdashline\Tstrut
        
        \texttt{\color{cnxtblue}ConvNeXt} & & & & & & & & & \\
        \hspace{0.3cm}\texttt{\color{orange}Conv2d} & \phantom{0}34 & 1 & 1 & 1 & $88\times88$ & (12, 16, 16, \phantom{0}68) & 2\,312 & 68 & 2\,380 \\
        \hspace{0.3cm}\texttt{\color{convred}Conv2d} & \phantom{0}34 & 3 & 1 & 4 & $120\times120$ & (12, 16, 16, 136) & 41\,616 & 136 & 41\,752 \\
        \hspace{0.3cm}\texttt{\color{convred}Conv2d} & 136 & 3 & 1 & 4 & $152\times152$ & (12, 16, 16, 136) & 166\,464 & 136 & 166\,600 \\
        \hspace{0.3cm}\texttt{\color{convblue}Conv2d} & 136 & 1 & 1 & 1 & $152\times152$ & (12, 16, 16, \phantom{0}68) & 9\,248 & 68 & 9\,316 \\
        \texttt{\color{grupurp}GRU} & & & & & & & & & \\
        \hspace{0.3cm}\texttt{\color{convblue}Conv2d} & 136 & 1 & 1 & 1 & $152\times152$ & (12, 16, 16, 136) & 18\,496 & 136 & 18\,632 \\
        \hspace{0.3cm}\texttt{\color{convblue}Conv2d} & 136 & 1 & 1 & 1 & $152\times152$ & (12, 16, 16, \phantom{0}68) & 9\,248 & 68 & 9\,316 \\

        \Tstrut
        \texttt{\color{convtbrown}ConvTrans2d} & \phantom{0}68 & 2 & 2 & 1 & $154\times154$ & (12, 32, 32, \phantom{0}68) & 18\,496 & 68 & 18\,476 \\
        \texttt{Concat} (2) & --- & --- & --- & --- & --- & (12, 32, 32, 136) & 0 & 0 & 0 \\
        \texttt{\color{cnxtblue}ConvNeXt} & & & & & & & & & \\
        \hspace{0.3cm}\texttt{\color{convred}Conv2d} & 136 & 3 & 1 & 2 & $154\times154$ & (12, 32, 32, 272) & 332\,928 & 272 & 333\,200 \\
        \hspace{0.3cm}\texttt{\color{convred}Conv2d} & 272 & 3 & 1 & 2 & $162\times162$ & (12, 32, 32, 272) & 665\,856 & 272 & 666\,128 \\
        \hspace{0.3cm}\texttt{\color{convblue}Conv2d} & 272 & 1 & 1 & 1 & $170\times170$ & (12, 32, 32, 136) & 36\,992 & 136 & 37\,128 \\
        \texttt{\color{grupurp}GRU} & & & & & & & & & \\
        \hspace{0.3cm}\texttt{\color{convblue}Conv2d} & 272 & 1 & 1 & 1 & $170\times170$ & (12, 32, 32, 272) & 73\,984 & 272 & 74\,256 \\
        \hspace{0.3cm}\texttt{\color{convblue}Conv2d} & 272 & 1 & 1 & 1 & $170\times170$ & (12, 32, 32, 136) & 36\,992 & 136 & 37\,128 \\

        \Tstrut
        \texttt{\color{convtbrown}ConvTrans2d} & 136 & 2 & 2 & 1 & $171\times171$ & (12, 64, 64, 136) & 73\,984 & 136 & 74\,120 \\
        \texttt{Concat} (1) & --- & --- & --- & --- & --- & (12, 64, 64, 272) & 0 & 0 & 0 \\
        \texttt{\color{cnxtblue}ConvNeXt} & & & & & & & & & \\
        \hspace{0.3cm}\texttt{\color{orange}Conv2d} & 272 & 1 & 1 & 1 & $171\times171$ & (12, 64, 64, 136) & 36\,992 & 136 & 37\,128 \\
        \hspace{0.3cm}\texttt{\color{convred}Conv2d} & 272 & 3 & 1 & 1 & $173\times173$ & (12, 64, 64, 544) & 1\,331\,712 & 544 & 1\,332\,256 \\
        \hspace{0.3cm}\texttt{\color{convred}Conv2d} & 544 & 3 & 1 & 1 & $175\times175$ & (12, 64, 64, 544) & 2\,663\,424 & 544 & 2\,663\,968 \\
        \hspace{0.3cm}\texttt{\color{convblue}Conv2d} & 544 & 1 & 1 & 1 & $175\times175$ & (12, 64, 64, 136) & 73\,984 & 136 & 74\,120 \\
        \texttt{\color{grupurp}GRU} & & & & & & & & & \\
        \hspace{0.3cm}\texttt{\color{convblue}Conv2d} & 272 & 1 & 1 & 1 & $175\times175$ & (12, 64, 64, 272) & 73\,984 & 272 & 74\,256 \\
        \hspace{0.3cm}\texttt{\color{convblue}Conv2d} & 272 & 1 & 1 & 1 & $175\times175$ & (12, 64, 64, 136) & 36\,992 & 136 & 37\,128 \\
        \texttt{\color{convblue}Conv2d} & 136 & 1 & 1 & 1 & $175\times175$ & (12, 64, 64, \phantom{0}14) & 1\,904 & 14 & 1\,918 \\
        \midrule
         & & & & & & & 9\,816\,752 & 6\,066 & 9\,822\,818\\
        \bottomrule 
    \end{tabularx}
\end{table}
\subsubsection{Residual Prediction}\label{sec:methods.model.residual_prediction}
We switch to a residual prediction approach both for the full predictive step and within each ConvNeXt block. \revB{The ConvNeXt block \cite{liu2022convnet} is designed to minimize compute, while maintaining performance. 
It introduces an inverted channel-bottleneck where the kernel size is reduced to $k=1$. This saves parameters and compute, because channel depth is only processed with a $1\times1$ spatial filter. As shown in \autoref{fig:model}, though, we modify the original ConvNeXt block from \citeA{liu2022convnet} by implementing a kernel size of $k=3$ and employing a two-stage convolution as done in \citeA{weyn2021sub}.}

Predicting \revB{residuals, that is, changes over a time step}, instead of \revB{full  values}, is similar to the discretization of time derivatives when solving partial or ordinary differential equations, and has been used successfully in previous DLWP models  \cite{hu2022swinvrnn,keisler2022forecasting,lam2022graphcast,pathak2022fourcastnet}.

\subsubsection{Inverting the Ordering of Channel Depth}
The standard U-Net for semantic segmentation \cite{ronneberger2015u} and its successors \cite{zhou2018unet++,huang2020unet} employ relatively few channels on the highest level and successively double the channel depth, while halving the spatial resolution in each deeper layer. This ordering is useful in image segmentation tasks, where deeper channels are required to create increasingly abstract filters to identify semantic features and express complex objects. In weather prediction, however, we find it is better to devote more capacity to the layers in the first level, where a wide variety of fine grained weather phenomena must be captured. Deeper layers at coarser resolution, on the other hand, need only encode larger scale atmospheric motions, which can be adequately represented with comparably fewer channels.

Thus, we invert the channel order, employing 136, 68, and 34 channels in each convolution on the first, second, and third layer, respectively (cf., \autoref{fig:model}). While this modification improves the model performance significantly, it also increases the computational burden, since more computations and data processing are required to evaluate the additional convolutions at fine spatial resolution. Tests which preserved the total number of trainable parameters, but completely eliminated the deeper layers in the U-Net gave worse results, demonstrating that the longer-range connections and richer latent space structures enabled by the full U-Net architecture remain important.

\subsubsection{Recurrent Modules}\label{sec:methods.model.recurrent_modules}
The vanilla U-Net is a feed-forward network, which treats successive inputs independently even if the data represents a continuous sequence over time. Feed-forward networks do not have any memory capacity. They do not maintain an internal state between time steps. To exploit information from previous latent states, we include a gated recurrent unit (GRU) \cite{cho2014learning} at the end of each decoder block, \revB{implemented as a convolutional GRU \cite{ballas2015delving} with $1\times 1$ spatial convolutions. GRUs use a hidden latent state that accumulates information over time to influence the current forecast step.} We chose GRUs over LSTMs \cite{hochreiter1997long} since we re-initialize the recurrent data over each 24-h cycle, and therefore do not require forget-gates (as confirmed experimentally, not shown).

\subsubsection{Miscellaneous Modifications}
Several other components of the original \citeA{weyn2021sub} model were modified based on recent results from deep learning research: The capped leaky ReLU was replaced by capped GELU activation functions \cite{hendrycks2016gaussian};\footnote{\revB{Gaussian error linear units (GELUs) are characterized by a smooth derivative that facilitates the optimization of deep learning models. We cap the maximum of the linear GELU part to 10 in order to prevent exploding activities in long rollouts.}} upsampling was changed from \revA{nearest-neighbor sampling  (knn-sampling with $k=1$)} to a transposed convolution; finally, the pairs of two successive convolutions were replaced at each encoder and decoder level in the U-Net by a modified ConvNeXt block \cite{liu2022convnet}, as visualized in \autoref{fig:model}.

\subsubsection{Time Stepping Scheme}
Similarly to \citeA{weyn2021sub}, we apply a two-in-two-out mapping with a temporal resolution twice as fine as the actual time step. For example, two atmospheric states \SI{3}{h} apart (each consisting of seven prognostic, along with three prescribed fields) are concatenated and input to the model, which generates a new pair of states, each characterising the atmosphere \SI{6}{h} later in time. This strategy is observed to stabilize and accelerate the training, since the model receives additional information about the atmosphere's rate of change and only has to be called half as often.

The frequency spectrum of atmospheric kinetic energy has a strong peak at \SI{24}{h} because many circulations are modulated by solar heating. We therefore evaluate the training loss function as the mean squared error over a 24-h period. Tests in which the MSE was evaluated over multi-day periods tended to result in a model that gradually approached climatology over many recursive steps.

\begin{figure}
    \centering
    \includegraphics[width=\textwidth]{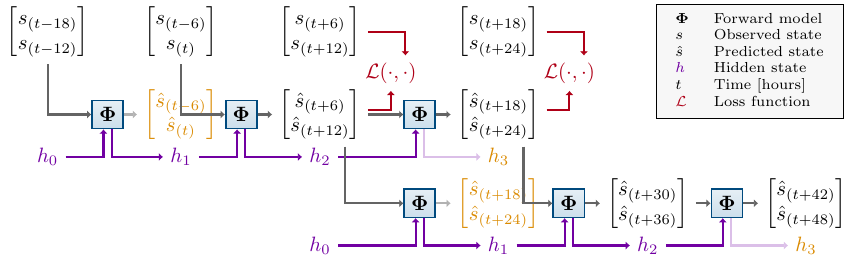}
    \caption{Two time-level input-output scheme with GRU for training and inference assuming \SI{6}{h} time resolution. The output from the preliminary initialization step (\revB{in orange}) is discarded, but the hidden state $h_1$ is generated and used in the first model step.  The hidden state $h_3$ (\revB{in orange}) at the end of the \SI{24}{h} forecast is discarded as the GRU will be re-initialized for the next recursive inference step (lowest row). For training (top right), the loss function is computed from the four forecast times spanning a \SI{24}{h} period at \SI{6}{h} resolution, as indicated in red.}
    \label{fig:forward_schematic}
\end{figure}

Training our model only over one daily cycle  does mean that the recurrent states of the GRUs are not optimized for long rollouts.
To prevent the explosion of recurrent states when generating long multi-day forecasts, we re-initialize the recurrent states every \SI{24}{h} as illustrated in \autoref{fig:forward_schematic} for a 12-h time step with \SI{6}{h} resolution. For training or for the first step in a long forecast rollout, the model predicts $\smash{[\hat{s}_{(t+6)}, \hat{s}_{(t+12)}]}$ from initial data $\smash{[{s}_{(t-6)}, {s}_{(t)}]}$, and then in the subsequent step uses $\smash{[\hat{s}_{(t+6)}, \hat{s}_{(t+12)}]}$ to predict $\smash{[\hat{s}_{(t+18)}, \hat{s}_{(t+24)}]}$. But before this, the hidden states for the GRUs are initialized in a preliminary step by calling the model once with the state pair $\smash{[s_{(t-18)}, s_{(t-12)}]}$ and a hidden state $h_0$ initialized with zeros. The resulting forecast for $\smash{[\hat{s}_{(t-6)}, \hat{s}_{(t)}]}$ is discarded, but the hidden state $h_1$ is supplied to the GRU and paired with the actual initial data $\smash{[{s}_{(t-6)}, {s}_{(t)}]}$ for the first step of the model. As shown by the bottom row in \autoref{fig:forward_schematic}, in a forecast rollout, the next day's prediction begins by re-initializing the GRU starting with forecast values from one time step earlier and $h_0$ set to zero to obtain $h_1$. Note that since the GRU is re-initialized every day, there would be five model steps per day when using a \SI{6}{h} time step (with \SI{3}{h} data resolution).

\subsubsection{Training}
Our best performing DLWP-HPX model, described above, has \SI{9.8}{M} parameters that are trained for 300 epochs (equivalent to 931,199 update steps) over eight days on four NVIDIA A100 GPUs with \SI{80}{GB} VRAM each. A batch size of eight per GPU is chosen, effectively resulting in an overall batch size of 32. We combine the Adam optimizer \cite{kingma2014adam} with a cosine annealing learning rate scheduler \cite{loshchilovsgdr}, setting the initial learning rate to \SI{2e-4}{} and gradually refining it to zero. To stabilize the training, we clip the gradients to the current learning rate, which we observe to be particularly beneficial for large recurrent models.

\subsection{\revA{The Receptive Field}}\label{sec:methods.receptive_field}

Several leading DLWP models \cite{pathak2022fourcastnet,hu2022swinvrnn,bi2023pangu,chen2023fengwu} are based on Vision Transformers (ViTs) \cite{dosovitskiy2020image}, which were originally developed to account for non-local \revA{relationships} in images; effectively working on patch embeddings.
ViTs are successors of Transformers \cite{vaswani2017attention}, which were introduced to efficiently accommodate very non-local \revA{relationships} in natural language processing (NLP), where no fixed upper bound exists on the distance between words that may interact to change the meaning of a sentence. In contrast to ViTs, we use a U-Net to emphasize local \revA{atmospheric} interactions, \revA{nevertheless each step of our model samples from a very large receptive field.} \revB{(The ``receptive field" is the set of grid cells the model accesses when generating output for a specific target pixel.)} 

There is a strong physical constraint on the locality of atmospheric interactions, \revA{which is that {\it no atmospheric disturbances travel faster than the speed of sound,}} roughly 
\SI{300}{m/s}. 
Sound waves are not meteorologically significant, \revA{and are not represented in the data used to train ML weather models.}
A better measure of the speed of the fastest moving signals of meteorological importance is the transport by the strongest jet-stream winds, \revA{which could transport a passive tracer} at roughly \SI{100}{m/s}, or about \SI{4300}{km} in \SI{12}{h}. 

The pair of 2$\times$2 average poolings and the dilations in the second and third levels of our U-Net architecture (\autoref{fig:model}) substantially widen the receptive field that potentially influences the solution at a given point after each forward step of our model. 
Neglecting influences from special points at the corners of the twelve basic HEALPix faces, the receptive field at each stage of the neural network is listed in \autoref{app:tab:layer_parameters} and grows to a 175$\times$175 patch of cells after the last $3\times3$ convolution in the decoder. 

The diagonal distance between adjacent points on our 3$\times$3 stencil (dark blue patch in \autoref{fig:healpix}) on a HPX64 mesh is approximately \SI{110}{km}. Thus, the receptive field for one step of our full HPX64 model is a patch exceeding \SI{18900}{km} on each side, which is large enough to include all points influenced by sound wave propagation over a \SI{12}{h} time step, and far more than would be required to contain the fastest moving meteorologically \revA{significant signals present in the ERA5 training data. In particular, at} every step, our HPX64 forecast at a given point is influenced by a set of surrounding points containing roughly 70\% of all the cells covering the globe.

\section{Results}

In the following, we first evaluate key variables in our model over a 14-day forecast lead time, which is slightly longer than the period over which knowledge of the initial atmospheric conditions gives these single deterministic forecasts some predictive skill. We compare our best model with the ECMWF S2S forecasts and with our previous \citeA{weyn2021sub} results. We then document the successive improvements that our changes in model architecture have on the RMSE and ACC scores for $Z_{500}$.
Next, we examine the ability of the model to distinguish between the amplitudes of the daily $T_{2m}$ ranges in tropical forests, in deserts, and over the ocean. Finally, we examine the behavior of the simulations over sub-seasonal (eight-week) and one-year free running rollouts.

\subsection{Quantitative Performance Through 14-Day Forecast Lead Time}\label{sec:results.2week}

To compare our model with the results from \citeA{weyn2021sub} and to state-of-the-art NWP from ECWMF, we compute both root mean squared error (RMSE) between observations and model predictions and anomaly correlation coefficient (ACC) scores with respect to the ERA5 climatology. Both metrics are compared on a $1^{\circ}\times 1^{\circ}$ lat-lon mesh and weighted by latitude, requiring us to project our DLWP-HPX and \citeA{weyn2021sub} forecasts from the HEALPix and cubed sphere meshes onto the lat-lon grid. Because our ultimate focus is on sub-seasonal and seasonal forecasting, we compare against ECMWF's integrated forecasting system for sub-seasonal forecasts (IFS S2S), which were initialized bi-weekly on Mondays and Thursdays and stepped forward at about \SI{16}{km} effective resolution for the first 15 days (then doubling to \SI{32}{km}).\revA{\footnote{\url{https://confluence.ecmwf.int/display/S2S/ECMWF+model+description}}} For comparison with \citeA{weyn2021sub}, our test set focuses on the years 2017 and 2018. In this and all the following cases, except a few simulations in our ablation study, computations are performed at HPX64 and \SI{3}{h} resolution (corresponding to \SI{6}{h} time steps).

\revB{To further compare our model with a state-of-the-art DLWP model, we include $Z_{500}$ scores for GraphCast, retrieved from the interactive WeatherBench2 \cite{rasp2023weatherbench} homepage.\footnote{\url{https://sites.research.google/weatherbench/deterministic-scores/}} In contrast to the others, GraphCast scores are computed on its native $0.25^{\circ}\times 0.25^{\circ}$ grid and for 2018 only, since the model was trained on data including 2017.} Key parameter attributes of the model from \citeA{weyn2021sub}, IFS S2S\revB{, GraphCast}, and our HPX64 model are listed in \autoref{tab:model_properties}.

 \revB{The GraphCast-WeatherBench2-RMSE scores at $T_{850}$, and particularly at $T_{2m}$, are difficult to compare with those from our model at early forecast lead times because differences in resolution and grid structure influence the representation of the topography and coastlines. Therefore we only plot GraphCast scores at $Z_{500}$. As previously documented, the RMSE and ACC of GraphCast temperature forecasts at $0.25^{\circ}\times 0.25^{\circ}$ resolution, are somewhat better than those from the IFS \cite{lam2022graphcast}.}
 
\begin{table}
    \centering
    \caption{Number of trainable parameters in millions, number of spherical shells of prognostic variables, horizontal resolution in degrees latitude, and temporal resolution ($\Delta_t$) of the models compared in \autoref{fig:rmse_acc_best}.}
    \label{tab:model_properties}
    \small
    \begin{tabularx}{0.7\textwidth}{>{\hsize=0.2\hsize}l >{\hsize=0.2\hsize}R >{\hsize=0.2\hsize}R >{\hsize=0.2\hsize}R >{\hsize=0.2\hsize}R}
        \toprule
        Model & Parameters & Spherical shells  & Resolution & $\Delta_t$\\
        \midrule
        Weyn 2021 & 2.7M & 6 & $1.4^\circ$ & \SI{6}{h} \\
        Our HPX64 & 9.8M &  7 & $1^\circ$ & \SI{3}{h} \\
        ECMWF & --- & $900+$ & $0.15^\circ$ & \SI{0.2}{h} \\
        \revB{GraphCast} & 21M & 227 & $0.25^\circ$ & \SI{6}{h} \\
        \bottomrule
    \end{tabularx}
\end{table}

\begin{figure}
    \centering
    \includegraphics[width=\textwidth]{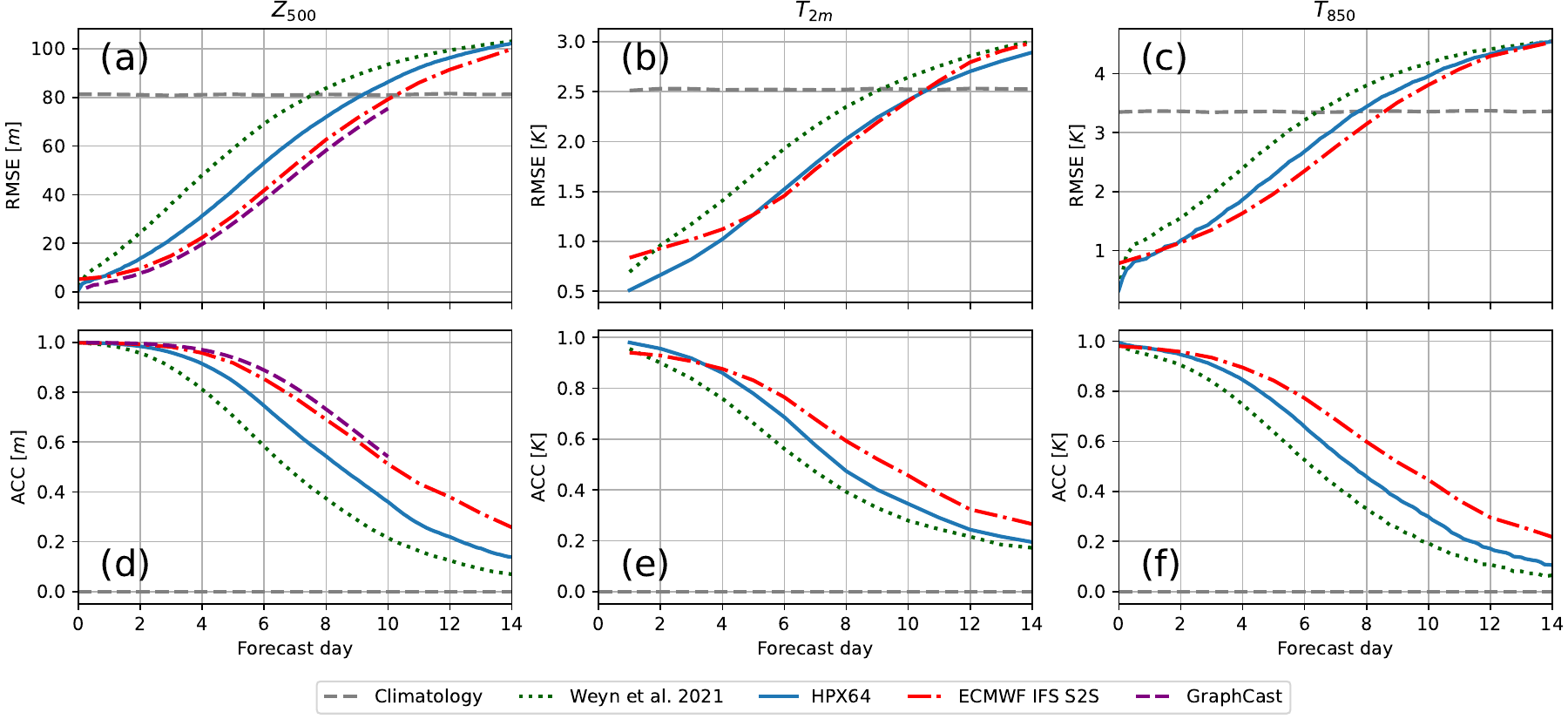}
    \caption{Comparison of the performance of the DLWP-HPX, Weyn et al. (2021), ECMWF IFS S2S\revB{, and GraphCast} models. \revB{GraphCast is averaged over 104 forecasts for 2018, while other forecasts are averaged over 204 forecasts from 2017 through 2018}. RMSE for (a) $Z_{500}$, (b) $T_{2m}$, and (c) $T_{850}$; climatology is indicated by the gray dashed line.  ACC for or (d) $Z_{500}$, (e) $T_{2m}$ and (f) $T_{850}$.}
    \label{fig:rmse_acc_best}
\end{figure}

As shown in \autoref{fig:rmse_acc_best}, the RMSE scores for $Z_{500}$, 24-hour-averaged $T_{2m}$ (because instantaneous $T_{2m}$ fields are not archived from the ECMWF S2S forecasts\footnote{\url{https://apps.ecmwf.int/datasets/data/s2s-realtime-daily-averaged-ecmf/levtype=sfc/type=cf/}}) and $T_{850}$ all improve substantially compared to \citeA{weyn2021sub}. Moreover, despite the small number of prognostic variables and coarse spatial resolution of our model, the RMSEs for $Z_{500}$ only lag the scores for ECMWF S2S and \revB{GraphCast} by about 1 day at one-week lead time. The HPX64 RMSE for $T_{850}$ shows a similar lag in skill compared to the IFS. As expected theoretically, the RMSE scores for all  models appear to be asymptotically approaching $\smash{\sqrt{2}}$ times climatology beyond two weeks when the skill of a single deterministic forecast drops toward zero. We present the comparison of 24-hour-averaged $T_{2m}$ between our model and IFS S2S for completeness, but it should be interpreted with caution. The re-gridding of both the IFS S2S and the HEALPix data to the $1^{\circ}\times 1^{\circ}$ lat-lon analysis grid introduces errors in the representation of coastlines and topography that significantly influence the surface temperature field. As a consequence, the RMSE values shown in \autoref{fig:rmse_acc_best} (b) are not representative of those in each model's native representation of the $T_{2m}$ field.

One additional issue that arises when plotting initial RMSE (and to a lesser extent ACC) for the ECMWF IFS S2S model is that, unlike our DLWP-HPX model, the IFS forecasts are not initialized with the ERA5 data. Thus, at very short forecast lead times, differences between the IFS initialization and the ERA5 data introduce apparent errors in the IFS forecast that are not representative of its actual performance. \citeA{lam2022graphcast} accounted for this in their comparison between the IFS and GraphCast, but it requires considerable extra computation. We are not claiming to outperform the IFS, so we simply suggest using caution when comparing errors between our models and the IFS at lead times less than 2 days.

ACC scores for $Z_{500}$, $T_{2m}$, and $T_{850}$ are also shown in \autoref{fig:rmse_acc_best}(d)--(f). As with RMSE, there is substantial improvement relative to both the previous model from \citeA{weyn2021sub} and the IFS S2S. In meteorological contexts, an ACC score of 0.6 is typically considered the lower limit of practical skill. The scores from our HEALPix model cross this threshold at about 7.5 days for $Z_{500}$ and 6.5 days for $T_{850}$, both of which are about 1.5 day sooner than the respective results for the IFS S2S \revB{and for the GraphCast $Z_{500}$ forecast}. Numerical comparisons of the model RMSE and ACC scores averaged over the same 208 forecasts used to plot \autoref{fig:rmse_acc_best} are given for 3-day and 5-day lead times in \autoref{tab:rmse_acc}.

\begin{figure}
    \centering
    \includegraphics[width=\textwidth]{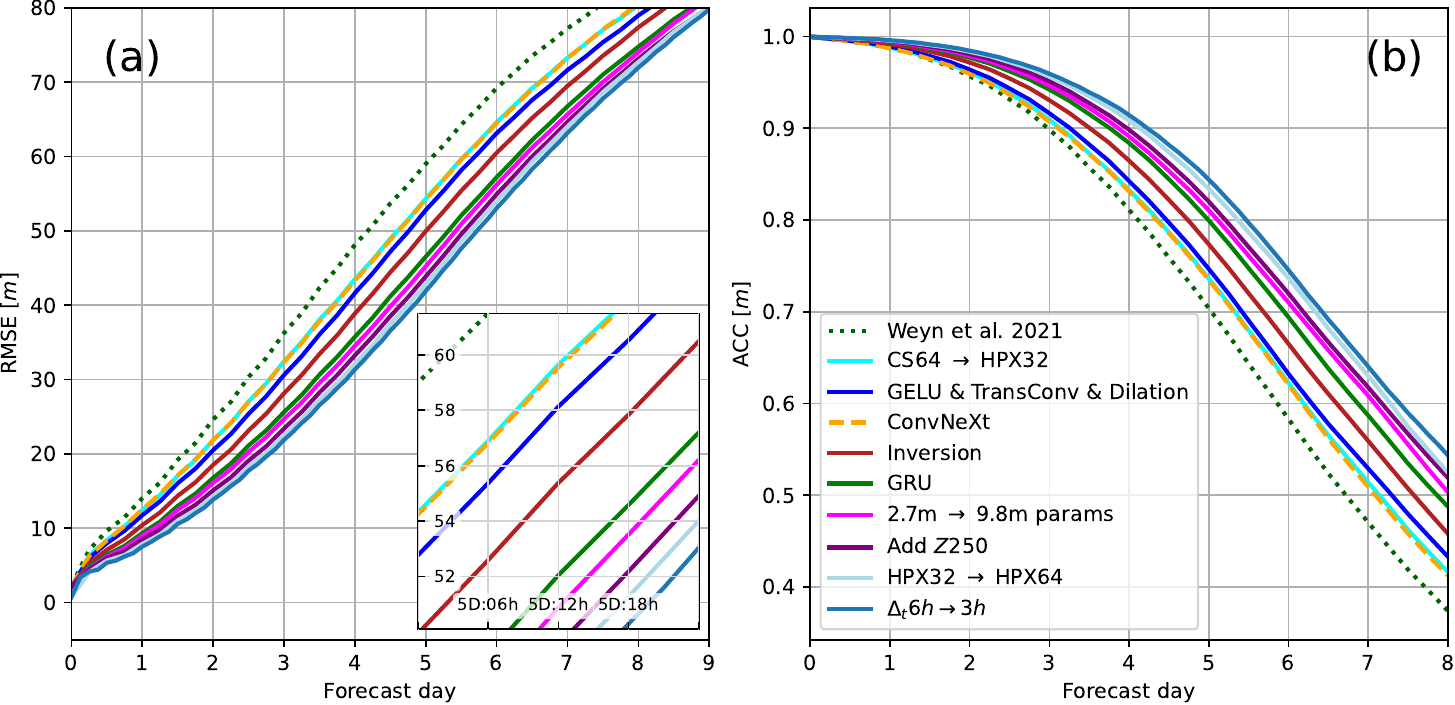}
    \caption{Impact of successive model improvements on the accuracy of $Z_{500}$ RMSE. 
    Each successive change builds on top of the previous architecture, adding the modification indicated in the legend: (a) RMSE, (b) ACC. Inset in (a) provides a magnified view of the error growth between 5 and 6 forecast days.}
    \label{fig:rmse_acc_advancements}
\end{figure}

The relative importance of the various improvements in model architecture between \citeA{weyn2021sub} and our best DLWP-HPX model is illustrated for the $Z_{500}$ field in \autoref{fig:rmse_acc_advancements}. The total number of trainable parameters is held constant at roughly $2.7\times 10^6$  over the first five sets of changes. The RMSE rises to \SI{50}{m} around 4.2 days in \citeA{weyn2021sub} (dark green dotted curve); replacing the $64\times 64$ cubed sphere by a HPX32 grid (aqua curve) delays the error growth by about 0.5 day despite the associated 50\% reduction in total grid points. There is also a similar substantial improvement in the ACC. Continuing with the HPX32 mesh, we replace the capped ReLU by a capped GELU activation function, replace knn-interpolation by strided transposed convolution, and introduce dilated convolutions in the two lower levels of the U-Net (as detailed in \autoref{fig:model}); this yields the modest but distinct improvements shown by the dark-blue curves.

Next, we replace the pairs of convolutions in each level of the encoder and decoder by a ConvNeXt block \revB{with kernel size $k=3$} (dashed tan curve). This actually produces a slight degradation in performance, but in other configurations closer to our final model, the ConvNeXt block does improve the performance, and importantly, it also reduces the memory footprint by about $25\%$ at a constant parameter count.  A further significant improvement is obtained by inverting the standard U-Net progression in channel depth to have the most channels at the highest spatial resolution and the fewest at the lowest resolution (dark red curve). The final significant improvement in the 2.7-million parameter model is obtained by adding recurrence in the form of GRU cells in the decoder (green curve).

After adding the GRU cells, the rise of the RMSE to \SI{50}{m} is delayed to about 5.3 days and the drop of the ACC below 0.6 to roughly 6.8 days. The next series of changes produces successive small improvements that push these values out to about 5.7 days for RMSE and 7.4 days for ACC. These improvements, as sequentially plotted in \autoref{fig:rmse_acc_advancements}, are: increasing the number of trainable parameters to $9.8\times 10^6$, adding the $Z_{250}$ field, increasing the horizontal resolution to HPX64 (which is more important for ACC than RMSE particularly on $T_{2m}$), and decreasing the time resolution to \SI{3}{h}. Benefits from the use of 3-h time resolution were only obtained if the model was configured with the GRUs.

\begin{table}
    \centering
    \caption{Root mean squared errors (RMSE) and anomaly correlation coefficient (ACC) scores for Weyn et al. (2021) (W21), our HPX64, and ECMWF's IFS models, evaluated on geopotential at \SI{500}{hPa} ($Z_{500}$), temperature \SI{2}{m} above ground ($T_{2m}$), and temperature at \SI{850}{hPa} ($T_{850}$) on lead times of 3 and 5 days.}
    \label{tab:rmse_acc}
    \begin{tabularx}{\textwidth}{llRRRRRRRRR}
        \toprule
         & & \multicolumn{3}{c}{$Z_{500}$} & \multicolumn{3}{c}{$T_{2m}$} & \multicolumn{3}{c}{$T_{850}$} \\
        \cmidrule(r){3-5}\cmidrule(lr){6-8}\cmidrule(l){9-11}
         & Lead time & W21 & HPX64 & IFS & W21 & HPX64 & IFS & W21 & HPX64 & IFS \\
        \midrule
        \multirow{2}{*}{\rotatebox[origin=c]{90}{\parbox[c]{0.6cm}{\centering \scriptsize{RMSE}}}} & 3 days & 36.26 & 21.88 & 14.91 & 1.17 & 0.82 & 1.02 & 1.95 & 1.49 & 1.35 \\
         & 5 days & 59.01 & 41.91 & 31.30 & 1.67 & 1.27 & 1.27 & 2.83 & 2.28 & 1.96 \\
        \hdashline\Tstrut
        \multirow{2}{*}{\rotatebox[origin=c]{90}{\parbox[c]{0.6cm}{\centering \scriptsize{ACC}}}} & 3 days & 0.90 & 0.96 & 0.98 & 0.84 & 0.92 & 0.91 & 0.84 & 0.91 & 0.94 \\
         & 5 days & 0.70 & 0.84 & 0.92 & 0.66 & 0.78 & 0.83 & 0.64 & 0.76 & 0.84 \\
        \bottomrule
    \end{tabularx}
\end{table}

The single most effective modification in the preceding set of successive improvements is the migration from the cubed sphere to the HEALPix mesh, even though the $64\times 64$ cubed sphere has twice the total number of grid-points as the HPX32 mesh. \revB{A} likely explanation for the superiority of the HEALPix mesh is not simply that it is a more uniform covering of the globe than that provided by the cubed sphere, but that \revB{it allows us to train a single set of location-invariant kernels for use over the entire globe. Note that} east and west have the same orientation in every HEALPix cell; we refer to this property as ``east to the right.'' In particular, the center and the east and west corners of each HEALPix cell are all at the same latitude. (A similar relationship holds in the north-south direction for meridians passing through those cells lying equatorward of the maximum north-south extent of the four equatorial faces in \autoref{fig:healpix} (a).)
Thus, on the HEALPix mesh, eastward motion at all points and at all latitudes would be in the same direction across the diamond-shaped $3\times 3$ stencil in \autoref{fig:healpix} (c). 
In contrast, at any point on either of the polar faces on the cubed sphere, east could map to any of four directions along the axes of the $3\times3$ convolutional stencil, depending on its longitude, as visualized in \ref{app:sec:deep_learnin_on_the_healpix}.

\revB{Since most large-scale weather systems move in a generally eastward direction in mid and high latitudes, we believe the ``east-to-the-right'' property allows a fixed number of kernel elements to more efficiently produce the required set of flow evolutions in the latent layers. This is because we can train one set of kernels for use everywhere on the HEALPix mesh instead of training separate sets of kernels for the equatorial and for the polar faces on the cubed sphere \cite{weyn2021sub}. A HEALPix model with the same total number of trainable parameters as the cubed sphere model can, therefore, employ twice as many trainable elements within each kernel.} 

\subsection{Eliminating the Need for Boundary-Layer Parameterizations}

Accurate forecasts of surface temperatures in NWP models rely on the empirical parameterization of multi-scale processes near the Earth's surface in the atmospheric boundary layer (ABL). The bottom of the ABL includes the roughness layer (2--5 times the height of roughness elements such as vegetation), and the surface layer (often \SI{10}{}--\SI{100}{m} deep), where shear-driven turbulence dominates generation by convection.  The depth of the full ABL, where larger-scale eddies and circulations communicate the processes in the surface layer to the free atmosphere, can vary from $O$(100)\,\unit{\meter} in calm stable nighttime conditions to several kilometers during the day over deserts.  

\begin{figure}
    \centering
    \includegraphics[width=0.7\textwidth]{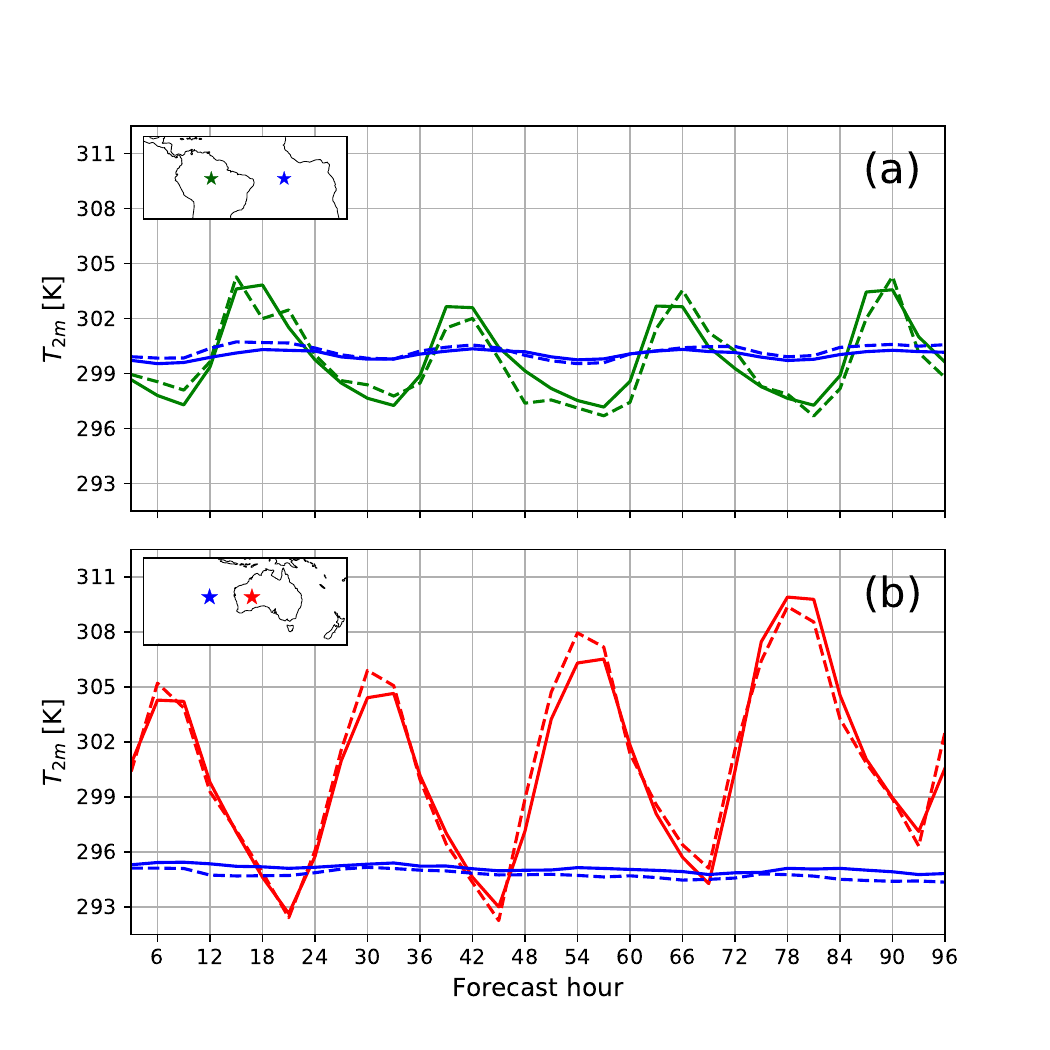}
    \caption{HPX64 simulation of the diurnal cycle of $T_{2m}$ (solid curves) at the four locations shown in the insets starting from 00 UTC on 12 March 2018. ERA5 values for the same $1^{\circ}\times 1^{\circ}$ lat-lon cell are shown as dashed lines.  Values are plotted every \SI{3}{h}.}
    \label{fig:diurnal_cycle}
\end{figure}

No effort is made to explicitly account for ABL processes in our model; the $T_{2m}$ field is treated the same as the other six prognostic fields. The same CNN kernels are employed everywhere over the globe on the HEALPix mesh; the only data that might distinguish one location from another are the land-sea mask, the terrain elevation, and the TOA solar forcing; neither longitude nor latitude are provided. Yet our model does a good job of capturing the diurnal cycle in multi-day forecasts over very different surfaces.  \autoref{fig:diurnal_cycle} shows the diurnal cycle in $T_{2m}$ at locations over the Amazon forest, the Australian desert, and two adjacent oceans over a 4-day simulation starting at 00 UTC on 12 March 2018.  

Compared to over land, the diurnal $T_{2m}$ variations are modest over the oceans, and they are well captured by our model. The land-sea mask is undoubtedly important in distinguishing the ocean locations from those over land. More interestingly, the model does an excellent job of capturing the large diurnal temperature range over the Australian desert, while correctly generating a much lower amplitude signal over the Amazon. The prognostic field that has most likely facilitated this distinction is $TCWV$, which is significantly higher over the Amazon than over the Australian desert. The model also captures the 4-day trend for increasing temperatures over Australia, which is linked to the evolution of larger-scale weather systems. Overall, the ability of the model to capture the diurnal $T_{2m}$ cycle with just seven prognostic fields, without any special treatment of the ABL, and without geo-specific inputs such as latitude and longitude is suggestive of the power and potential of DLWP-HPX.

\subsection{\revA{Iterative Rollouts Over Subseasonal to Annual Time Scales}}

\revA{There are three time scales of primary interest for global atmospheric simulations: medium-range weather forecasting for lead times of up to two weeks, sub-seasonal and seasonal forecasts for lead times up to 6--9 months, and climate simulations over periods of tens to hundreds of years.  Our focus is on the sub-seasonal to seasonal time scale; therefore, in this section we examine the model's performance in iterative rollouts over periods up to one year.}  

\begin{figure}
    \centering
    \includegraphics[width=\textwidth]{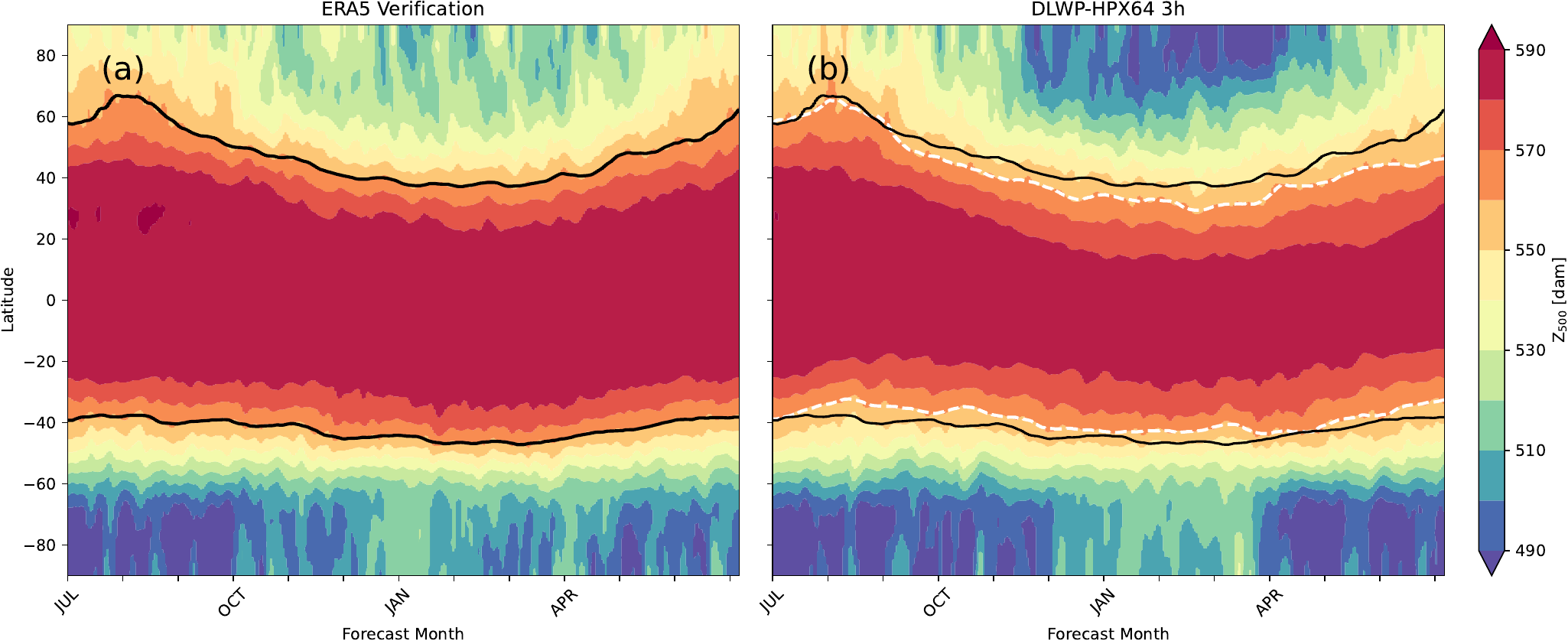}
    \caption{Zonally averaged three-day mean of $Z_{500}$ plotted as a function of time and latitude for one year beginning on July 1 2017 for: (a) the ERA5 reanalysis, and (b) a recursive one-year rollout of the DLWP-HPX model. Also shown are 15-day averaged values of the \SI{5600}{m}  contour of $Z_{500}$ for the ERA5 data (black lines) the DLWP-HPX simulation (white dashed lines).}
    \label{fig:365-days}
\end{figure}

\revA{To investigate the stability and drift in model simulations over a full annual cycle}, we initialize it using  ERA5 data for 00 UTC on 1 June 2017 (together with the 21 UTC fields on 31 May).  Using 6-h time steps (with 3-h time resolution), we perform 1460 iterations to generate a 365-day \revA{simulation}. The three-day running mean of $Z_{500}$, averaged around each latitude, is plotted as a function of latitude and time  in \autoref{fig:365-days}, along with the corresponding averages from the ERA5 data. Despite being trained to minimize RMSE over a single day and not enforcing any physical constraints, the DLWP-HPX simulation responds to the TOA solar forcing to generate the annual cycle reasonably well.

\begin{figure}
    \centering
    \includegraphics[width=\textwidth]{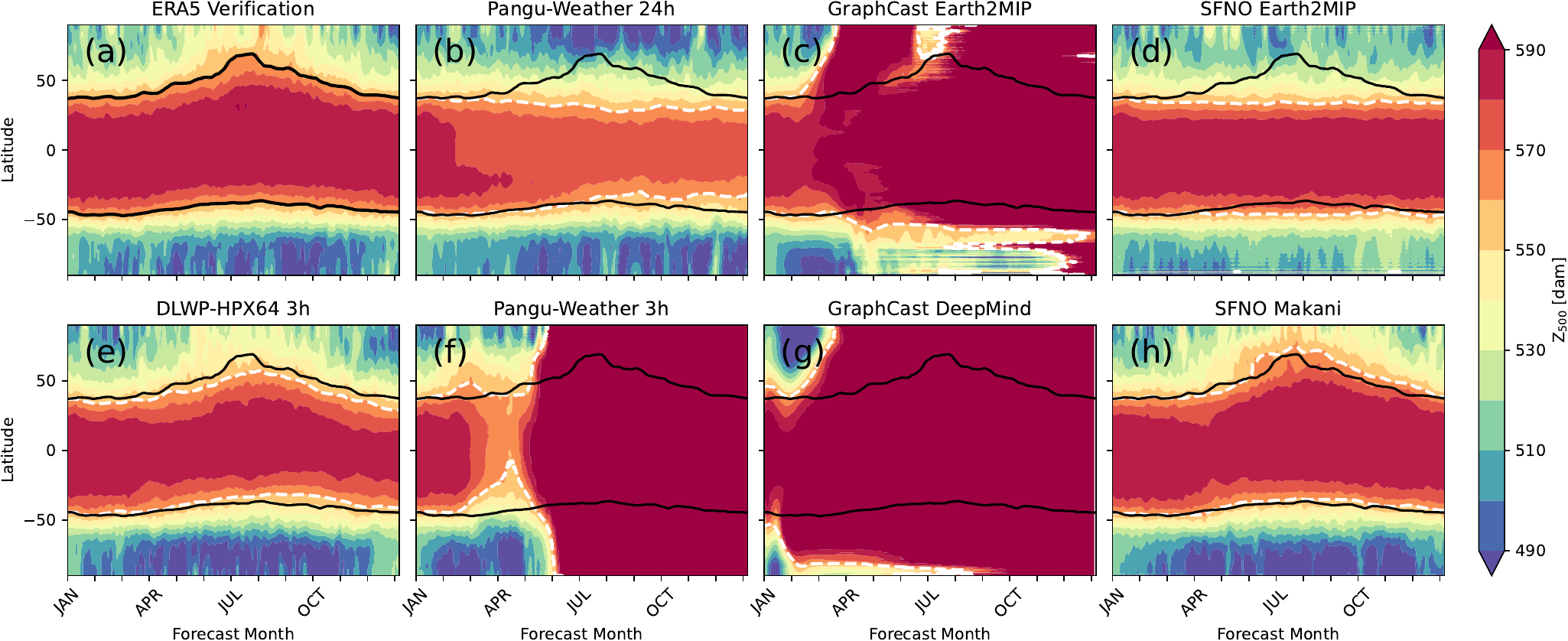}
    \caption{Zonally averaged three-day mean of $Z_{500}$ plotted as a function of time and latitude: (a) for ERA5 reanalysis, (b)-(h) for recursive one-year simulations for each model as identified in the titles, initialized on January 2, 2018. 
    Also shown are 15-day averaged values of the \SI{5600}{m}  contour of $Z_{500}$ for the ERA5 data (black lines) each model simulation (white dashed lines).}
    \label{fig:365-days_sota}
\end{figure}

One region where the errors are significant is the arctic. About 5 months into the simulation, the simulated heights in the arctic region drop as much as 60 m below those in the reanalysis during the boreal winter. In contrast, at 5--8-month lead times, the heights in the antarctic region increase to approximately correct values in the austral summer. \revA{The asymmetry between the response in arctic and antarctic flips if the one-year rollout begins six months later. When the simulation is initialized on January 2, 2018, the heights in the arctic during boreal winter are approximately correct, while those in the antarctic are too cold (\autoref{fig:365-days_sota}d).}

There is also a long-term drift toward lower heights in the subtropics and mid-latitudes, creating a roughly \SI{30}{m} loss in $Z_{500}$  by the end of the 1-year forecast.\footnote{\SI{30}{m} deviation amounts to 0.5\% of the full $Z_{500}$ value and to 8.7\% of the $Z_{500}$ standard deviation (computed from the reanalysis data of the forecasted period).} Climate models are tuned to avoid long-term drift in the predicted fields, but operational NWP models are not so tuned. For example, significant model biases that grow over a time scale of several weeks are removed to create sub-seasonal ECMWF IFS S2S forecasts \cite{Vitart2004,Weigel2008}. To facilitate comparison of model drift with the ERA5 reanalysis, the pair of black lines in both panels show the 15-day mean of the zonally averaged 560-dam $Z_{500}$ contours in the northern and southern hemisphere. The white lines in \autoref{fig:365-days}b show the corresponding 560-dam $Z_{500}$ contours for the DLWP-HPX simulation. The drift toward lower heights starts to become evident after two months in the northern hemisphere and continues to grow slowly for the remainder of the year.  Differences show up earlier in the southern hemisphere, but the average drift is smaller and even disappears at a few times later in the year. \revA{As will be discussed in a forthcoming paper, both the errors near the poles and the drift in the tropics in $Z_{500}$ can be corrected by incorporating SST forecasts from a coupled atmosphere-ocean model.}

\revA{The performance of three additional state-of-the-art DLWP models is compared with our model using this same metric in \autoref{fig:365-days_sota}, which shows the  evolution of zonally averaged $Z_{500}$ heights over a one-year rollout beginning January 2, 2018. This year is part of the test set for all of the models: our DLWP-HPX, Pangu-Weather, GraphCast, and FourCastNetv2 based on spherical Fourier neural operators (SFNO) \cite{bonev2023sfno}. Details about the code used to generate these rollouts can be found in the Open Research Section.} 

\revA{The Pangu-Weather model does not include solar forcing, and therefore, it does not follow the annual cycle.  When stepped forward with a 24-h time step (\autoref{fig:365-days_sota}b), significant drift is apparent after about 1.5 months, which grows through the year without pushing the simulation into grossly unrealistic states. Based on the discussion of Extended Data, Fig.~7a in \cite{bi2023pangu}, one would not expect good performance from Pangu-Weather if rolled out with a 3-h time step, and indeed the 3-h rollout starts to produce significant errors after 1.5 months and generates completely unrealistic results after about 5 months (\autoref{fig:365-days_sota}f). We nevertheless, show its performance to contrast it with our 3-h-time-resolution rollout (\autoref{fig:365-days_sota}e).} 

\revA{The version of GraphCast from NVIDIA's Earth2MIP gives reasonable results for just the first 1.5 months (\autoref{fig:365-days_sota}c), while that from DeepMind goes bad after a couple weeks (\autoref{fig:365-days_sota}g). The SFNO Earth2MIP model (FourCastNetv2-small) shows essentially no drift over a full year (\autoref{fig:365-days_sota}d), \revB{but it does not follow the annual cycle because it neglects changes in solar forcing.}  Some artifacts (horizontal stripes) are visible near the south pole within a month and at the north pole much later in the simulation. In contrast, the SFNO Makani model (\autoref{fig:365-days_sota}h) includes solar zenith angle as an input field, and it does follow the annual cycle reasonably well. On balance, the performance of the SFNO Makani model is roughly similar to our DLWP-HPX model; it has larger errors near the poles, but less drift in the tropics. }

In an ablation study (not shown), we investigated the effect of the top-of atmosphere solar forcing input on the 365-day DLWP-HPX rollout by training a model that did not receive solar forcing input. In that case, the model still generated a stable forecast over the entire rollout period, but did not produce the full annual cycle. Interestingly, that simulation did roughly approximate the transition from summer into a perpetual autumn.

\begin{figure}
    \centering
    \includegraphics[width=0.8\textwidth]{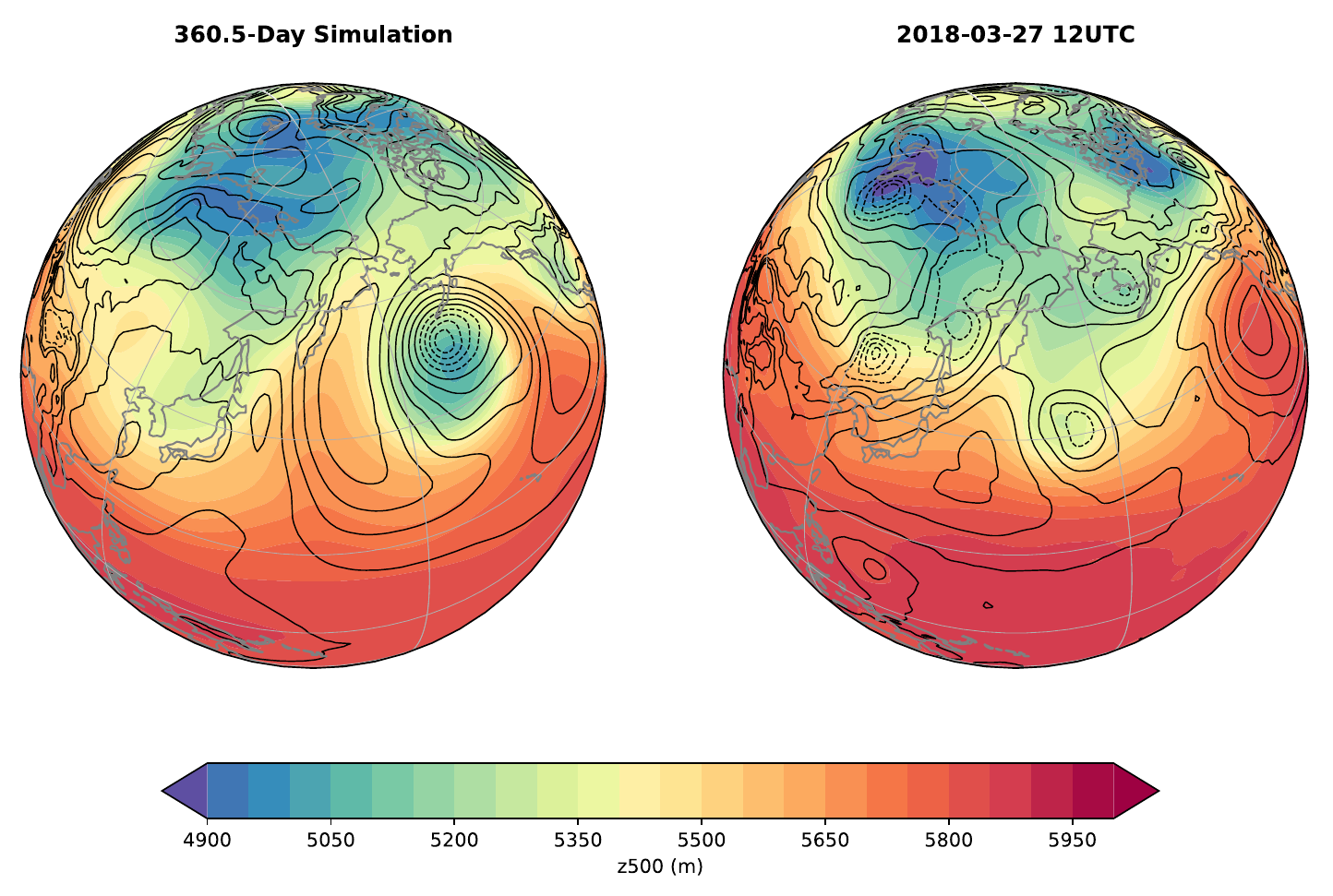}
    \caption{$Z_{500}$ (color fill: \SI{50}{dam} contour interval) and $Z_{1000}$ (black contours: \SI{40}{m} interval) for a free-running 360.5-day simulation \revB{(1442 autoregressive steps)} and the corresponding ERA5 reanalysis for 00 UTC on 27 March 2018. Dashed black lines indicate values of $Z_{1000}\le\SI{40}{m}$ (corresponding to sea-level pressures less than roughly \SI{1008}{hPa}).}
    \label{fig:360.5_day_map}
\end{figure}

One qualitative way to appreciate the \revB{ability of our model to retain realistic weather patterns in a 1442-step rollout} is illustrated by comparing a 360.5 day simulation initialized on 1 April 2017 (with 3-h resolution) and the corresponding 27 March 2018 reanalysis in \autoref{fig:360.5_day_map}.
The roughly one-year lead time is well beyond the limits of atmospheric predictability, so there is no reason to expect a close match between simulation and reanalysis. The 360.5-day simulation time was chosen to display the simulated strong low-pressure center in the northeastern Pacific. The intensity of the system is typical for strong systems in our simulation, but its lowest $Z_{1000}$ heights are about \SI{40}{m} higher than those in the strongest systems periodically appearing in the ERA5 reanalysis. Lower-amplitude signals also appear in the $Z_{1000}$ field, which is somewhat less than \SI{50}{m} too low in the tropics. On balance, the overall character of this late-March weather pattern is quite plausible. \revB{In some models that use latitude-longitude meshes, obvious errors at the poles can show up in as little as 10 autoregressive steps \cite[Fig.~4]{bonev2023sfno}. As evident in \autoref{fig:360.5_day_map}, no artifacts are apparent in the vicinity of the North Pole after 1442 autoregressive steps.} 

\begin{figure}
    \centering
    \includegraphics[width=0.6\textwidth]{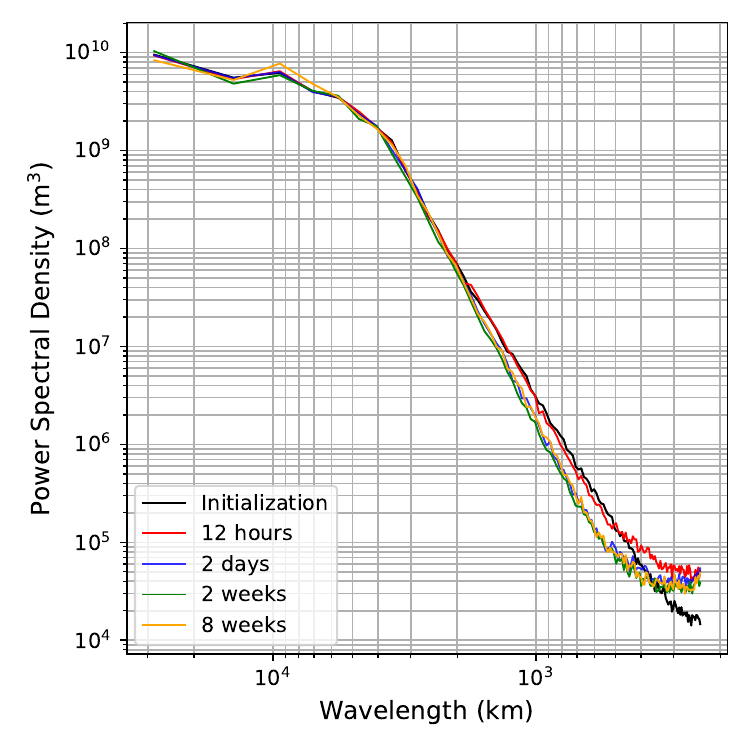}
    \caption{One dimensional power spectral density of the $Z_{500}$ field around the $45^{\circ}$ N latitude, averaged over 208 bi-weekly forecasts from 2017-2018 at: initialization (black), and at forecast  lead times of \SI{12}{h}, \SI{2}{d},  2, and 8 weeks.}
    \label{fig:Z500_spectral}
\end{figure}

A more quantitative assessment of any tendency of our model to distort the atmospheric state by damping or amplifying mid-latitude perturbations at different wavelengths is provided by the plots of the $Z_{500}$ power spectral density around $\SI{45}{^\circ N}$ in \autoref{fig:Z500_spectral}. These spectra are averaged over 208 biweekly forecasts from the 2017-2018 test set \revB{for which the RMSE and ACC were plotted in \autoref{fig:rmse_acc_best}}. The initial spectrum in black represents the average state of the atmosphere in the ERA5 reanalysis. 

Twelve hours (2 recursive steps) after initialization there is very little change in the spectra for wavelengths $\lambda$ longer than \SI{500}{km} (roughly 5 grid intervals), but the power in the shorter waves is amplified. Over the next \SI{36}{h}, there is a gradual reduction in the amplitude at wavelengths $\lambda < \SI{1800}{km}$ to yield a spectrum that is somewhat damped over the interval $380 < \lambda < \SI{1800}{km}$ and amplified at the shortest wavelengths. Surprisingly, the spectral distribution at two days remains essentially unchanged \revB{throughout the subsequent autogressive rollout} at least out to sub-seasonal-forecast lead times of eight weeks \revB{(244 steps)}, which is consistent with the impression obtained by examining images such as those in \autoref{fig:360.5_day_map}. 

\revB{What does the deviation of the spectral power from the correct ERA5 curve imply about the ability of the model to approximate a true atmospheric state? As part of the answer, important quantitative points of reference are the RMSE and ACC errors for $Z_{500}$ at day 2 plotted in \autoref{fig:rmse_acc_best}. The day-2 global RMSE error over the same set of forecasts and verifications for which spectra are plotted in \autoref{fig:Z500_spectral} is about \SI{17}{m}; the ACC is negligibly different from the correct value of 1.0. Theses values represent upper bounds on the 2-day forecast error that might be produced exclusively by the spectral distortion of the $Z_{500}$ field because other factors also contribute to the RMSE and ACC error, such as incorrectly approximating the speed and direction at which features propagate. Of course there is no deterministic predictability at 8-week forecast lead times, but since the 8-week spectrum in \autoref{fig:Z500_spectral} is essentially identical to that at 2 days, the DLWP-HPX 8-week forecasts need not be farther from some realizable atmospheric state than what is suggested by the modest 2-day $Z_{500}$ errors in \autoref{fig:rmse_acc_best}a,d.}

\section{Conclusion}

We have presented an improved CNN-based DLWP-HPX model that stably forecasts atmospheric evolution over a full one-year cycle using a very limited set of prognostic variables. The number of actual degrees of freedom characterising predictable atmospheric states at forecast lead times beyond 3--5 days is not known, but is far less than the total number of prognostic variables carried at every grid cell in state-of-the-art NWP models. Here, we have demonstrated that realistic atmospheric simulations can be performed using just seven prognostic variables above each cell  on a HEALPix mesh with \SI{110}{km} between the nodes.

The HEALPix mesh \cite{gorski2005healpix} has been used in astronomy for almost two decades, but has previously seen very little use in atmospheric science. The mesh covers the sphere with a hierarchical grid of equal-area cells uniformly spaced along circles at constant latitudes. A particularly important advantage of the HEALPix mesh for weather forecasting with CNNs is that it is an ``east to the right'' mesh, i.e., east has the same orientation in every HEALPix cell. Weather systems tend to travel west-to-east in mid- and high-latitudes and both east-to-west (tropical cyclones) or west-to-east (Madden-Julian Oscillation, convectively coupled Kelvin waves) in the tropics. The kernel weights in our convolutional stencils can more economically learn this behavior than on our previous cubed sphere mesh in which the eastward orientation across the stencil varies with longitude, particularly on the polar faces. \revB{More importantly, because all cells have the same east-to-the-right orientation, we do not need to train separate sets of convolution filters for the equatorial and polar regions. Thus, a HEALPix model with the same total number of trainable parameters as a cubed sphere can employ twice as many filter weights as that used for cubed sphere. Although switching from a cubed sphere mesh with $64\times 64$ cells on each of the six faces to a HEALPix mesh with $32\times 32$ cells on each of the 12 faces reduces the total number of grid points covering the sphere by half, it increases the time over which  the $Z_{500}$ RMSE remains below \SI{40}{m} by almost 1/2 day at a 4-day} forecast lead time (\autoref{fig:rmse_acc_advancements}). 

Two other significant improvements to our model architecture were obtained by adding recursion via GRUs and by inverting the standard way channel depth is refined at deeper layers in the U-Net. In contrast to the original U-Net architecture \citeA{ronneberger2015u}, our channel depth halves instead of doubles as the spatial resolution is also halved in each successively deeper U-Net layer. This allows the model to devote more trainable parameters to describing the wide variety of fine-scale weather patterns while using comparatively fewer parameters to describe the simpler set of global weather patterns. Although this modification pushes the U-Net toward the basic ResNet architecture \cite{he2016deep}, we find the deeper U-Net layers continue to provide significant skill to the forecasts.

Additional modest improvements were implemented by switching to the GELU activation function and to $2\times 2$ transposed strided convolutions when up-sampling; by increasing the total number of trainable parameters from \SI{2.7}{M} to \SI{9.8}{M}, adding the $Z_{250}$ field, increasing the resolution to HPX64, and increasing the time resolution to \SI{3}{h} (which gives us a \SI{6}{h} time step). The benefits of 3-h time resolution were only realized when the model included the GRUs. The 3-h time resolution gives a good forecast of the daily cycle of surface temperature, and the model also learns the difference in the range of that cycle between regions of tropical forest and desert without geo-specific input data. 

Finally, we replaced the pairs of successive convolutions in \citeA{weyn2020improving} with modified ConvNeXt blocks. The switch to the ConvNeXt blocks was only advantageous at higher resolutions, where in addition to improving accuracy, it reduced the memory footprint.

At one-week forecast lead time, the resulting model is roughly 1 day behind the ECMWF IFS S2S forecast error in $Z_{500}$ RMSE and 1.5 days behind in ACC. \revB{Our} statistics are worse than those for Pangu-Weather \cite{bi2023pangu} and GraphCast \cite{lam2022graphcast}, both of which provide $Z_{500}$ RMSE and ACC forecasts at $0.25^{\circ}\times 0.25^{\circ}$ resolution that are superior to the \revA{deterministic} ECMWF IFS high-resolution model averaged to the same $0.25^{\circ}\times 0.25^{\circ}$ grid. Despite having less accuracy in medium range forecasts, our model \revA{can be recursively stepped forward to generate better \SI{500}{hPa} forecasts over seasonal and one-year rollouts than GraphCast and Pangu-Weather. It is also superior to the SFNO version of FourCastNetv2 currently on NVIDIA Earth2MIP, though it behaves similarly to the recently checkpointed version of SFNO Makani. Realistic low pressure systems and upper-level trough and ridge patterns continue to be generated by our model at the end of the one-year rollout.}


Deep learning models for weather forecasting are evolving rapidly, with important advancements using a wide variety of architectures. \revB{A common methodology in atmospheric science research involves the investigation of some phenomena using a hierarchy of models with decreasing complexity, such as GCMs with full physics parameterizations, simpler nonlinear numerical models with minimal parameterizations, and linear models with analytic solutions.  Our DLWP-HPX model provides an example of what can be achieved when training a parsimonious model on a server with just 4 NVIDIA A100 GPUs.} It may be particularly useful for scientific investigations when it is advantageous to work with a minimal set of unknown variables to more concisely characterize sensitivities that might be revealed by techniques such as backpropagation with respect to loss functions customized for analysis, \revB{as opposed to model training \cite{Ebert-Uphoff__2021}. As an example, note that the large-scale structure of the atmosphere is represented in our deepest U-Net layer on each time step by 34 latent-state variables on a coarse-resolution (\SI{440}{km}) grid. This information is decoded during each time step, along with finer resolution latent-state data from the skip connections, to give the updated physical state of the global system. We are currently designing classifier modules configured as a follower network to receive this deep latent-state information to explore the low-frequency variability of the atmosphere.}

There are many avenues along which our DLPW-HPX model might be improved. One would be to adding additional prognostic fields while carefully examining the resulting performance. Another one would lie in refining the CNN architecture, where the choice of particular inductive biases may be crucial \cite{thuemmel2023inductive}. A related important aspect of improving the modelled processes might be to incorporate explicit physical constraints, yielding physics-informed differentiable artificial neural networks \cite{beucler2021enforcing,Shen:2023}. Other natural extensions of this work lie in examining the performance of the DLPW-HPX model in ensemble forecasts, which are crucial to sub-seasonal and seasonal prediction and to couple the atmospheric model with the ocean, thus moving toward a deep learning earth system model \cite{Bauer:2023}. \revA{Preliminary results suggest that coupling our model with a deep learning ocean model that predicts sea surface temperatures (which are not incorporated in the current model) stabilizes the simulations and removes all model drift in multi-decadal rollouts.}



\appendix

\section{Deep Learning on the HEALPix}\label{app:sec:deep_learnin_on_the_healpix}

\subsection{Seamless Evolution of Location Invariant Kernels}
The Hierarchical Equal Area isoLatitude Pixelization (HEALPix) is a partitioning of the sphere that has found wide application in astronomy since it was introduced by \citeA{gorski2005healpix}. It divides the sphere into 12 base faces that can be hierarchically subdivided into patches of equal size. A key property for training CNNs on this mesh is the isolatitudinal alignment, that is, patches are aligned along lines of latitude and each patch has the same orientation, which we describe as ``east to the right'' in \autoref{sec:results.2week}.

\begin{figure}
    \centering
    \includegraphics[width=\textwidth]{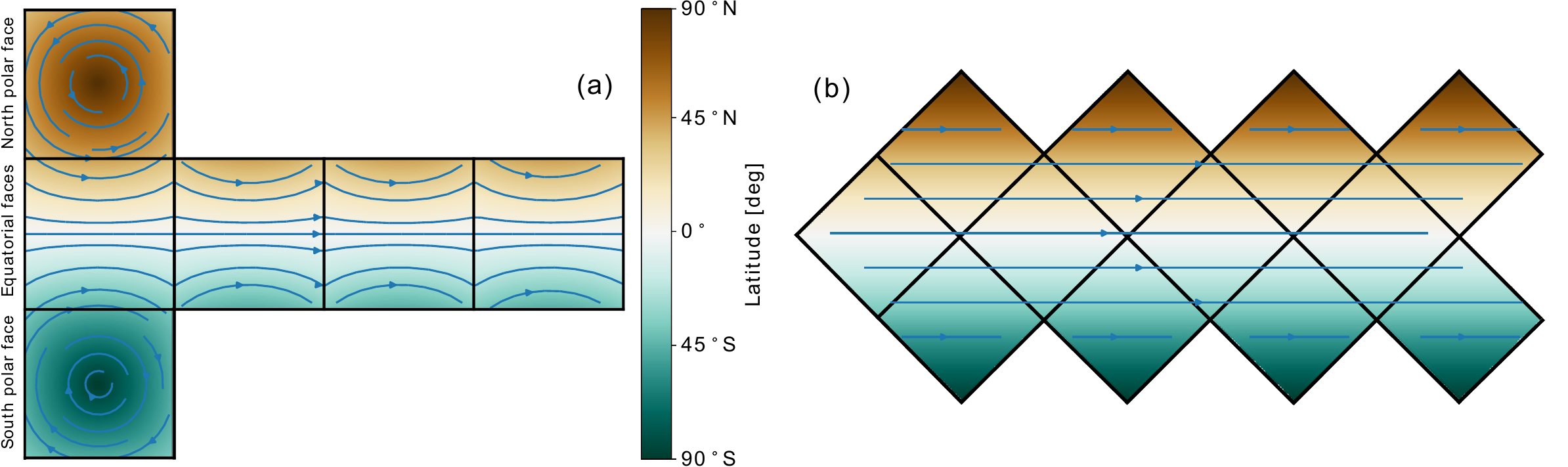}
    \caption{Lines of latitudes depicted as blue streamline arrows on the cubed sphere (a) and on the HEALPix (b). While the lines corresponding to constant eastward motion describe arcs of different radii on the cubed sphere mesh, the same motion translates to straight lines on the HEALPix mesh.}
    \label{fig:app:cubed_sphere_lines_of_latitude}
\end{figure}

To contrast and emphasize the difficulty that CNN kernels are facing on the cubed sphere mesh, we plot the lines of constant latitude on the six faces of the cubed sphere and on the twelve faces of the HEALPix in \autoref{fig:app:cubed_sphere_lines_of_latitude}. Except for the equator, all lines of constant latitude are bent on the cubed sphere, imposing challenges for a limited set of convolution kernels that must evolve location invariant pattern detectors and functions. For example, \revB{weather systems tend to migrate eastward in mid- and high-latitudes, and} the kernels need to learn a wider range of behaviors to propagate eastward motions at the top-left versus the \revB{bottom-right corners of the polar faces of the} cubed sphere face.

On the other hand, lines of constant latitude map to straight lines on the HEALPix mesh. This facilitates the formulation of location-invariant convolutional kernels for the propagation of weather systems, \revB{allowing the same set of kernels to be used over the entire globe. In contrast to the cubed sphere, it is not necessary to train separate sets of kernels for the equatorial and polar faces.  Therefore, without increasing the model's total number of trainable parameters, the convolutional kernels on the HEALPix mesh can accommodate more latent layers than on the cubed sphere.}

\subsection{Technical Implementation Details}
\begin{figure}
    \centering
    \includegraphics[width=\textwidth]{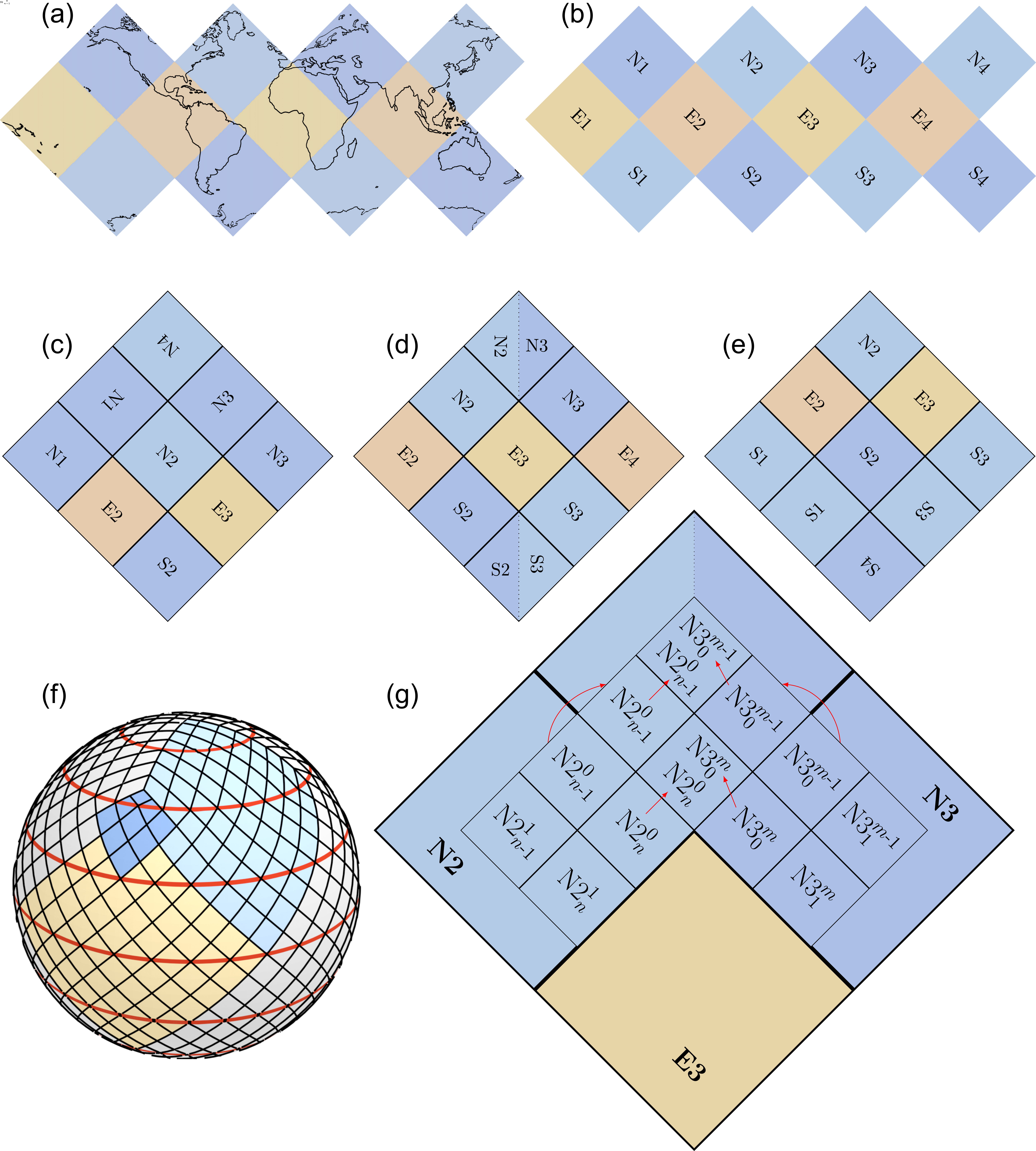}
    \caption{2D HEALPix face arrangement and padding. (a) depicts the distribution of coastlines over the twelve HEALPix faces. (b) enumerates the twelve faces of the HEALPix with each four faces on the northern and southern hemisphere and around the equator. (c), (d), and (e): Exemplary alignment and rotations of neighboring faces before applying the padding operation on northern (c), equatorial (d), and southern faces (e). (f) emphasizes the special corner case, which is detailed in (g) to visualize the padding. \revB{The missing corner pixel is filled by averaging the two values from the adjacent cells (row and column indices of each cell displayed as super- and subscripts, respectively).}}
    \label{fig:app:hpx_faces}
\end{figure}
Since deep learning libraries are optimized for image processing tasks, we consider each of the HEALPix's 12 base faces as a regular two-dimensional tensor, i.e., we interpret the sphere as a composition of twelve images (cf. \autoref{fig:healpix} and \autoref{fig:app:hpx_faces}).

To simulate the spatial propagation of dynamics beyond individual faces, such that weather patterns can evolve globally on the sphere, we implement custom padding operations to concatenate the relevant information of all neighboring faces to each respective face of interest.

\autoref{fig:app:hpx_faces} showcases our planet's coastlines projected on the HEALPix faces in (a) and outlines the spatial organization of the twelve faces in (b). The arrangement of neighboring faces is exemplarily detailed for the northern (N) and southern (S) hemisphere, as well as for the equatorial faces (E). To simulate the neighborhood of, say, face \texttt{E3}, the face \texttt{N2} must be concatenated to the left of \texttt{E3}, while face \texttt{S3} is concatenated to the right. On the northern and southern hemispheres, neighboring faces are partially required to be rotated, as indicated in \autoref{fig:app:hpx_faces} (c), (d), and (e).

A particular case occurs in the north and south corners of the tropical faces, where no natural neighbor exists---cf. \autoref{fig:healpix} and \autoref{fig:app:hpx_faces} (f) for an illustration. To simulate the ninth neighbor of the respective corner, we interpolate the values from the according faces on the northern/southern hemisphere, by simply averaging the two corresponding values and writing the result in the simulated neighboring face. For example, to simulate the top left neighboring face of \texttt{E3}, we average the respective values from \texttt{N2} and \texttt{N3}, as detailed by the straight red arrows in \autoref{fig:app:hpx_faces} (g). Values that do not lie on the main diagonal of the simulated face are not required to be interpolated, but are copied from the adjacent faces instead, denoted by the curved red arrows in \autoref{fig:app:hpx_faces} (g). The exemplary corner padding shows the case for the application of a $3\times3$ kernel with dilation of 1 or 2. Note that a $5\times5$ kernel could be applied in the same way. Importantly, the padding should not extend one neighboring face, which depends on the resolution of the HEALPix mesh and the configuration of the applied convolution (kernel size and dilation). Otherwise, a hierarchy of padding operations would be required to be implemented and considered.

%



%
%

\section*{Open Research Section}

Instructions for training, and a trained model for inference, are available at \url{https://github.com/CognitiveModeling/dlwp-hpx/}.  In addition, PyTorch code for training the DLWP-HPX model  is available in the repository at \url{https://github.com/NVIDIA/modulus/tree/main/examples/weather/dlwp_healpix}.
All spherical shells of data from ERA5 (Hersbach et al., 2020) were downloaded from Copernicus, where variables on various constant pressure levels, such as $Z_{500}$ or $T_{850}$, and variables on single levels, such as $T_{2m}$ or $TCWV$, are hosted open to the public, available at \url{https://cds.climate.copernicus.eu/cdsapp#!/dataset/reanalysis-era5-pressure-levels?tab=form} and \url{https://cds.climate.copernicus.eu/cdsapp#!/dataset/reanalysis-era5-single-levels?tab=overview}.

{To generate 1-year rollouts for Pangu-Weather, GraphCast, and FourCastNet2 (SFNO), as plotted in \autoref{fig:365-days_sota}, we considered the respective public repositories with the pretrained model weights. More concretely, we generated the SFNO Earth2MIP (\texttt{fcnv2\_sm}) and GraphCast Earth2MIP (\texttt{graphcast}) forecasts with NVIDIA's earth2mip package,\footnote{\url{https://github.com/NVIDIA/earth2mip}} specifically developing a custom script for long rollouts.\footnote{\url{https://github.com/NVIDIA/earth2mip/blob/main/examples/utils/workflows/1_year_run.py}} Checkpoints for the \texttt{SFNO Makani} forecast may be found in the NVIDIA NGC catalog.\footnote{\url{https://catalog.ngc.nvidia.com/orgs/nvidia/teams/modulus/models/sfno_73ch_small}} Interestingly, the original \texttt{GraphCast DeepMind} code base\footnote{\url{https://github.com/google-deepmind/graphcast}} produced slightly different results and saturated even faster than the Earth2MIP version, which might result from different random seeds. For the DeepMind version of GraphCast, we downloaded the model weights\footnote{\url{https://storage.googleapis.com/dm_graphcast/params/GraphCast\%20-\%20ERA5\%201979-2017\%20-\%20resolution\%200.25\%20-\%20pressure\%20levels\%2037\%20-\%20mesh\%202to6\%20-\%20precipitation\%20input\%20and\%20output.npz}} provided through their repository. Pangu-Weather forecasts in \SI{24}{h} and \SI{3}{h} resolution (with respective checkpoint files for the \SI{24}{h}\footnote{\url{https://drive.google.com/file/d/1lweQlxcn9fG0zKNW8ne1Khr9ehRTI6HP/view}} and \SI{3}{h}\footnote{\url{https://drive.google.com/file/d/1EdoLlAXqE9iZLt9Ej9i-JW9LTJ9Jtewt/view}} models) were generated by using the original repository.\footnote{\url{https://github.com/198808xc/Pangu-Weather}}}

\acknowledgments
We would like to thank Mauro Bisson from NVIDIA Corp. for providing optimized CUDA kernels for the HEALPix padding implementation, and Jonathan Weyn who previously implemented a code base on which this work was built. \revA{We thank Peter D\"uben\revB{, Imme Ebert-Uphoff,} and a \revB{third} anonymous reviewer for encouraging us to generate and compare the 1-year rollouts for other state-of-the-art DLWP methods and for other valuable suggestions.} This work received funding from Deutsche Forschungsgemeinschaft (DFG, German Research Foundation) under Germany's Excellence Strategy EXC 2064 – 390727645 and from the Office of Naval Research under grants N0014-21-1-2827 and N00014-22-1-2807. We thank the Deutscher Akademischer Austauschdienst (DAAD, German Academic Exchange Service) as well as the International Max Planck Research School for Intelligent Systems (IMPRS-IS) for supporting Matthias Karlbauer. Nathaniel was supported by a National Defense Science and Engineering Graduate Fellowship. We are grateful to NVIDIA and Stan Posey for the donation of A100 GPU cards. This research was additionally supported by a grant from the NVIDIA Applied Research Accelerator Program and utilized an NVIDIA DGX-100 Workstation. Moreover, this work benefited substantially from the barrier-free high quality ERA5 dataset provided by the ECMWF.

\section*{Author Roles}

Matthias implemented model, training and evaluation routines in \texttt{PyTorch}, as well as the HEALPix-related projection scripts under consideration of the \texttt{healpy} package, and drafted the manuscript together with Dale who supervised this project closely and who also made the model schematic in \autoref{fig:model}. Nathaniel was involved in discussions about model evolution and code structures and generated \autoref{fig:diurnal_cycle}, \autoref{fig:365-days}, and \autoref{fig:Z500_spectral}. Raul was involved in model discussions and generated \autoref{fig:360.5_day_map}. Thorsten helped with implementing the distributed PyTorch pipeline for multi-GPU training and with accelerating the process pipeline. \revA{Noah Brenowitz and Boris Bonev generated the 365-days rollouts with the Earth2MIP and Makani packages for SFNO and GraphCast}. Martin co-supervised this project and helped with proofreading and writing.

%
%

\bibliography{references}

\begin{thebibliography}{}

\bibitem [\protect \citeauthoryear {%
Ballas%
, Yao%
, Pal%
\BCBL {}\ \BBA {} Courville%
}{%
Ballas%
\ \protect \BOthers {.}}{%
{\protect \APACyear {2015}}%
}]{%
ballas2015delving}
\APACinsertmetastar {%
ballas2015delving}%
\begin{APACrefauthors}%
Ballas, N.%
, Yao, L.%
, Pal, C.%
\BCBL {}\ \BBA {} Courville, A.%
\end{APACrefauthors}%
\unskip\
\newblock
\APACrefYearMonthDay{2015}{}{}.
\newblock
{\BBOQ}\APACrefatitle {Delving deeper into convolutional networks for learning
  video representations} {Delving deeper into convolutional networks for
  learning video representations}.{\BBCQ}
\newblock
\APACjournalVolNumPages{arXiv preprint arXiv:1511.06432}{}{}{}.
\PrintBackRefs{\CurrentBib}

\bibitem [\protect \citeauthoryear {%
Battaglia%
\ \protect \BOthers {.}}{%
Battaglia%
\ \protect \BOthers {.}}{%
{\protect \APACyear {2018}}%
}]{%
battaglia2018relational}
\APACinsertmetastar {%
battaglia2018relational}%
\begin{APACrefauthors}%
Battaglia, P\BPBI W.%
, Hamrick, J\BPBI B.%
, Bapst, V.%
, Sanchez-Gonzalez, A.%
, Zambaldi, V.%
, Malinowski, M.%
\BDBL {}others%
\end{APACrefauthors}%
\unskip\
\newblock
\APACrefYearMonthDay{2018}{}{}.
\newblock
{\BBOQ}\APACrefatitle {Relational inductive biases, deep learning, and graph
  networks} {Relational inductive biases, deep learning, and graph
  networks}.{\BBCQ}
\newblock
\APACjournalVolNumPages{arXiv preprint arXiv:1806.01261}{}{}{}.
\PrintBackRefs{\CurrentBib}

\bibitem [\protect \citeauthoryear {%
Bauer%
\ \protect \BOthers {.}}{%
Bauer%
\ \protect \BOthers {.}}{%
{\protect \APACyear {2023}}%
}]{%
Bauer:2023}
\APACinsertmetastar {%
Bauer:2023}%
\begin{APACrefauthors}%
Bauer, P.%
, Dueben, P.%
, Chantry, M.%
, Doblas-Reyes, F.%
, Hoefler, T.%
, McGovern, A.%
\BCBL {}\ \BBA {} Stevens, B.%
\end{APACrefauthors}%
\unskip\
\newblock
\APACrefYearMonthDay{2023}{}{}.
\newblock
{\BBOQ}\APACrefatitle {Deep learning and a changing economy in weather and
  climate prediction} {Deep learning and a changing economy in weather and
  climate prediction}.{\BBCQ}
\newblock
\APACjournalVolNumPages{Nature Reviews Earth \& Environment}{4}{8}{507--509}.
\newblock
\begin{APACrefURL} \url{https://doi.org/10.1038/s43017-023-00468-z}
  \end{APACrefURL}
\newblock
\begin{APACrefDOI} \doi{10.1038/s43017-023-00468-z} \end{APACrefDOI}
\PrintBackRefs{\CurrentBib}

\bibitem [\protect \citeauthoryear {%
Bauer%
, Thorpe%
\BCBL {}\ \BBA {} Brunet%
}{%
Bauer%
\ \protect \BOthers {.}}{%
{\protect \APACyear {2015}}%
}]{%
bauer2015quiet}
\APACinsertmetastar {%
bauer2015quiet}%
\begin{APACrefauthors}%
Bauer, P.%
, Thorpe, A.%
\BCBL {}\ \BBA {} Brunet, G.%
\end{APACrefauthors}%
\unskip\
\newblock
\APACrefYearMonthDay{2015}{}{}.
\newblock
{\BBOQ}\APACrefatitle {The quiet revolution of numerical weather prediction}
  {The quiet revolution of numerical weather prediction}.{\BBCQ}
\newblock
\APACjournalVolNumPages{Nature}{525}{7567}{47--55}.
\PrintBackRefs{\CurrentBib}

\bibitem [\protect \citeauthoryear {%
Benjamin%
\ \protect \BOthers {.}}{%
Benjamin%
\ \protect \BOthers {.}}{%
{\protect \APACyear {2019}}%
}]{%
benjamin2019100}
\APACinsertmetastar {%
benjamin2019100}%
\begin{APACrefauthors}%
Benjamin, S\BPBI G.%
, Brown, J\BPBI M.%
, Brunet, G.%
, Lynch, P.%
, Saito, K.%
\BCBL {}\ \BBA {} Schlatter, T\BPBI W.%
\end{APACrefauthors}%
\unskip\
\newblock
\APACrefYearMonthDay{2019}{}{}.
\newblock
{\BBOQ}\APACrefatitle {100 years of progress in forecasting and NWP
  applications} {100 years of progress in forecasting and nwp
  applications}.{\BBCQ}
\newblock
\APACjournalVolNumPages{Meteorological Monographs}{59}{}{13--1}.
\PrintBackRefs{\CurrentBib}

\bibitem [\protect \citeauthoryear {%
Beucler%
\ \protect \BOthers {.}}{%
Beucler%
\ \protect \BOthers {.}}{%
{\protect \APACyear {2021}}%
}]{%
beucler2021enforcing}
\APACinsertmetastar {%
beucler2021enforcing}%
\begin{APACrefauthors}%
Beucler, T.%
, Pritchard, M.%
, Rasp, S.%
, Ott, J.%
, Baldi, P.%
\BCBL {}\ \BBA {} Gentine, P.%
\end{APACrefauthors}%
\unskip\
\newblock
\APACrefYearMonthDay{2021}{}{}.
\newblock
{\BBOQ}\APACrefatitle {Enforcing analytic constraints in neural networks
  emulating physical systems} {Enforcing analytic constraints in neural
  networks emulating physical systems}.{\BBCQ}
\newblock
\APACjournalVolNumPages{Physical Review Letters}{126}{9}{098302}.
\PrintBackRefs{\CurrentBib}

\bibitem [\protect \citeauthoryear {%
Bi%
\ \protect \BOthers {.}}{%
Bi%
\ \protect \BOthers {.}}{%
{\protect \APACyear {2023}}%
}]{%
bi2023pangu}
\APACinsertmetastar {%
bi2023pangu}%
\begin{APACrefauthors}%
Bi, K.%
, Xie, L.%
, Zhang, H.%
, Chen, X.%
, Gu, X.%
\BCBL {}\ \BBA {} Tian, Q.%
\end{APACrefauthors}%
\unskip\
\newblock
\APACrefYearMonthDay{2023}{}{}.
\newblock
{\BBOQ}\APACrefatitle {Accurate medium-range global weather forecasting with 3D
  neural networks} {Accurate medium-range global weather forecasting with 3d
  neural networks}.{\BBCQ}
\newblock
\APACjournalVolNumPages{Nature}{}{}{}.
\newblock
\begin{APACrefDOI} \doi{doi.org/10.1038/s41586-023-06185-3} \end{APACrefDOI}
\PrintBackRefs{\CurrentBib}

\bibitem [\protect \citeauthoryear {%
Bonev%
\ \protect \BOthers {.}}{%
Bonev%
\ \protect \BOthers {.}}{%
{\protect \APACyear {2023}}%
}]{%
bonev2023sfno}
\APACinsertmetastar {%
bonev2023sfno}%
\begin{APACrefauthors}%
Bonev, B.%
, Kurth, T.%
, Hundt, C.%
, Pathak, J.%
, Baust, M.%
, Kashinath, K.%
\BCBL {}\ \BBA {} Anandkumar, A.%
\end{APACrefauthors}%
\unskip\
\newblock
\APACrefYearMonthDay{2023}{}{}.
\newblock
{\BBOQ}\APACrefatitle {Spherical Fourier Neural Operators: Learning Stable
  Dynamics on the Sphere} {Spherical fourier neural operators: Learning stable
  dynamics on the sphere}.{\BBCQ}
\newblock
\APACjournalVolNumPages{arXiv preprint arXiv:2306.03838}{}{}{}.
\PrintBackRefs{\CurrentBib}

\bibitem [\protect \citeauthoryear {%
Charney%
, Fj{\"{o}}rtoft%
\BCBL {}\ \BBA {} Neumann%
}{%
Charney%
\ \protect \BOthers {.}}{%
{\protect \APACyear {1950}}%
}]{%
Charney1950}
\APACinsertmetastar {%
Charney1950}%
\begin{APACrefauthors}%
Charney, J\BPBI G.%
, Fj{\"{o}}rtoft, R.%
\BCBL {}\ \BBA {} Neumann, J\BPBI V.%
\end{APACrefauthors}%
\unskip\
\newblock
\APACrefYearMonthDay{1950}{}{}.
\newblock
{\BBOQ}\APACrefatitle {{Numerical Integration of the Barotropic Vorticity
  Equation}} {{Numerical Integration of the Barotropic Vorticity
  Equation}}.{\BBCQ}
\newblock
\APACjournalVolNumPages{Tellus A}{2}{4}{}.
\PrintBackRefs{\CurrentBib}

\bibitem [\protect \citeauthoryear {%
Chen%
\ \protect \BOthers {.}}{%
Chen%
\ \protect \BOthers {.}}{%
{\protect \APACyear {2023}}%
}]{%
chen2023fengwu}
\APACinsertmetastar {%
chen2023fengwu}%
\begin{APACrefauthors}%
Chen, K.%
, Han, T.%
, Gong, J.%
, Bai, L.%
, Ling, F.%
, Luo, J\BHBI J.%
\BDBL {}Ouyang, W.%
\end{APACrefauthors}%
\unskip\
\newblock
\APACrefYearMonthDay{2023}{}{}.
\newblock
{\BBOQ}\APACrefatitle {FengWu: Pushing the Skillful Global Medium-range Weather
  Forecast beyond 10 Days Lead} {Fengwu: Pushing the skillful global
  medium-range weather forecast beyond 10 days lead}.{\BBCQ}
\newblock
\APACjournalVolNumPages{arXiv preprint arXiv:2304.02948}{}{}{}.
\PrintBackRefs{\CurrentBib}

\bibitem [\protect \citeauthoryear {%
Cho%
\ \protect \BOthers {.}}{%
Cho%
\ \protect \BOthers {.}}{%
{\protect \APACyear {2014}}%
}]{%
cho2014learning}
\APACinsertmetastar {%
cho2014learning}%
\begin{APACrefauthors}%
Cho, K.%
, van Merrienboer, B.%
, Gulcehre, C.%
, Bougares, F.%
, Schwenk, H.%
\BCBL {}\ \BBA {} Bengio, Y.%
\end{APACrefauthors}%
\unskip\
\newblock
\APACrefYearMonthDay{2014}{}{}.
\newblock
{\BBOQ}\APACrefatitle {Learning phrase representations using RNN
  encoder-decoder for statistical machine translation} {Learning phrase
  representations using rnn encoder-decoder for statistical machine
  translation}.{\BBCQ}
\newblock
\BIn{} \APACrefbtitle {Conference on Empirical Methods in Natural Language
  Processing (EMNLP 2014).} {Conference on empirical methods in natural
  language processing (emnlp 2014).}
\PrintBackRefs{\CurrentBib}

\bibitem [\protect \citeauthoryear {%
Dosovitskiy%
\ \protect \BOthers {.}}{%
Dosovitskiy%
\ \protect \BOthers {.}}{%
{\protect \APACyear {2020}}%
}]{%
dosovitskiy2020image}
\APACinsertmetastar {%
dosovitskiy2020image}%
\begin{APACrefauthors}%
Dosovitskiy, A.%
, Beyer, L.%
, Kolesnikov, A.%
, Weissenborn, D.%
, Zhai, X.%
, Unterthiner, T.%
\BDBL {}others%
\end{APACrefauthors}%
\unskip\
\newblock
\APACrefYearMonthDay{2020}{}{}.
\newblock
{\BBOQ}\APACrefatitle {An image is worth 16x16 words: Transformers for image
  recognition at scale} {An image is worth 16x16 words: Transformers for image
  recognition at scale}.{\BBCQ}
\newblock
\APACjournalVolNumPages{arXiv preprint arXiv:2010.11929}{}{}{}.
\PrintBackRefs{\CurrentBib}

\bibitem [\protect \citeauthoryear {%
Dueben%
\ \BBA {} Bauer%
}{%
Dueben%
\ \BBA {} Bauer%
}{%
{\protect \APACyear {2018}}%
}]{%
Dueben2018design}
\APACinsertmetastar {%
Dueben2018design}%
\begin{APACrefauthors}%
Dueben, P\BPBI D.%
\BCBT {}\ \BBA {} Bauer, P.%
\end{APACrefauthors}%
\unskip\
\newblock
\APACrefYearMonthDay{2018}{}{}.
\newblock
{\BBOQ}\APACrefatitle {{Challenges and design choices for global weather and
  climate models based on machine learning}} {{Challenges and design choices
  for global weather and climate models based on machine learning}}.{\BBCQ}
\newblock
\APACjournalVolNumPages{Geoscientific Model Development}{11}{10}{3999--4009}.
\PrintBackRefs{\CurrentBib}

\bibitem [\protect \citeauthoryear {%
Ebert-Uphoff%
\ \protect \BOthers {.}}{%
Ebert-Uphoff%
\ \protect \BOthers {.}}{%
{\protect \APACyear {2021}}%
}]{%
Ebert-Uphoff__2021}
\APACinsertmetastar {%
Ebert-Uphoff__2021}%
\begin{APACrefauthors}%
Ebert-Uphoff, I.%
, Lagerquist, R.%
, Hilburn, K.%
, Lee, Y.%
, Haynes, K.%
, Stock, J.%
\BDBL {}Stewart, J\BPBI Q.%
\end{APACrefauthors}%
\unskip\
\newblock
\APACrefYearMonthDay{2021}{}{}.
\newblock
{\BBOQ}\APACrefatitle {{CIRA} Guide to Custom Loss Functions for Neural
  Networks in Environmental Sciences -- Version 1} {{CIRA} guide to custom loss
  functions for neural networks in environmental sciences -- version 1}.{\BBCQ}
\newblock
\APACjournalVolNumPages{arXiv preprint arXiv:2106.09757}{}{}{}.
\PrintBackRefs{\CurrentBib}

\bibitem [\protect \citeauthoryear {%
Gori%
, Monfardini%
\BCBL {}\ \BBA {} Scarselli%
}{%
Gori%
\ \protect \BOthers {.}}{%
{\protect \APACyear {2005}}%
}]{%
gori2005new}
\APACinsertmetastar {%
gori2005new}%
\begin{APACrefauthors}%
Gori, M.%
, Monfardini, G.%
\BCBL {}\ \BBA {} Scarselli, F.%
\end{APACrefauthors}%
\unskip\
\newblock
\APACrefYearMonthDay{2005}{}{}.
\newblock
{\BBOQ}\APACrefatitle {A new model for learning in graph domains} {A new model
  for learning in graph domains}.{\BBCQ}
\newblock
\BIn{} \APACrefbtitle {Proceedings. 2005 IEEE International Joint Conference on
  Neural Networks, 2005.} {Proceedings. 2005 ieee international joint
  conference on neural networks, 2005.}\ (\BVOL~2, \BPGS\ 729--734).
\PrintBackRefs{\CurrentBib}

\bibitem [\protect \citeauthoryear {%
Gorski%
\ \protect \BOthers {.}}{%
Gorski%
\ \protect \BOthers {.}}{%
{\protect \APACyear {2005}}%
}]{%
gorski2005healpix}
\APACinsertmetastar {%
gorski2005healpix}%
\begin{APACrefauthors}%
Gorski, K\BPBI M.%
, Hivon, E.%
, Banday, A\BPBI J.%
, Wandelt, B\BPBI D.%
, Hansen, F\BPBI K.%
, Reinecke, M.%
\BCBL {}\ \BBA {} Bartelmann, M.%
\end{APACrefauthors}%
\unskip\
\newblock
\APACrefYearMonthDay{2005}{}{}.
\newblock
{\BBOQ}\APACrefatitle {HEALPix: A framework for high-resolution discretization
  and fast analysis of data distributed on the sphere} {Healpix: A framework
  for high-resolution discretization and fast analysis of data distributed on
  the sphere}.{\BBCQ}
\newblock
\APACjournalVolNumPages{The Astrophysical Journal}{622}{2}{759}.
\PrintBackRefs{\CurrentBib}

\bibitem [\protect \citeauthoryear {%
Guibas%
\ \protect \BOthers {.}}{%
Guibas%
\ \protect \BOthers {.}}{%
{\protect \APACyear {2021}}%
}]{%
guibas2021efficient}
\APACinsertmetastar {%
guibas2021efficient}%
\begin{APACrefauthors}%
Guibas, J.%
, Mardani, M.%
, Li, Z.%
, Tao, A.%
, Anandkumar, A.%
\BCBL {}\ \BBA {} Catanzaro, B.%
\end{APACrefauthors}%
\unskip\
\newblock
\APACrefYearMonthDay{2021}{}{}.
\newblock
{\BBOQ}\APACrefatitle {Efficient Token Mixing for Transformers via Adaptive
  Fourier Neural Operators} {Efficient token mixing for transformers via
  adaptive fourier neural operators}.{\BBCQ}
\newblock
\BIn{} \APACrefbtitle {International Conference on Learning Representations.}
  {International conference on learning representations.}
\PrintBackRefs{\CurrentBib}

\bibitem [\protect \citeauthoryear {%
He%
, Zhang%
, Ren%
\BCBL {}\ \BBA {} Sun%
}{%
He%
\ \protect \BOthers {.}}{%
{\protect \APACyear {2016}}%
}]{%
he2016deep}
\APACinsertmetastar {%
he2016deep}%
\begin{APACrefauthors}%
He, K.%
, Zhang, X.%
, Ren, S.%
\BCBL {}\ \BBA {} Sun, J.%
\end{APACrefauthors}%
\unskip\
\newblock
\APACrefYearMonthDay{2016}{}{}.
\newblock
{\BBOQ}\APACrefatitle {Deep residual learning for image recognition} {Deep
  residual learning for image recognition}.{\BBCQ}
\newblock
\BIn{} \APACrefbtitle {Proceedings of the IEEE conference on computer vision
  and pattern recognition} {Proceedings of the ieee conference on computer
  vision and pattern recognition}\ (\BPGS\ 770--778).
\PrintBackRefs{\CurrentBib}

\bibitem [\protect \citeauthoryear {%
Hendrycks%
\ \BBA {} Gimpel%
}{%
Hendrycks%
\ \BBA {} Gimpel%
}{%
{\protect \APACyear {2016}}%
}]{%
hendrycks2016gaussian}
\APACinsertmetastar {%
hendrycks2016gaussian}%
\begin{APACrefauthors}%
Hendrycks, D.%
\BCBT {}\ \BBA {} Gimpel, K.%
\end{APACrefauthors}%
\unskip\
\newblock
\APACrefYearMonthDay{2016}{}{}.
\newblock
{\BBOQ}\APACrefatitle {Gaussian error linear units (gelus)} {Gaussian error
  linear units (gelus)}.{\BBCQ}
\newblock
\APACjournalVolNumPages{arXiv preprint arXiv:1606.08415}{}{}{}.
\PrintBackRefs{\CurrentBib}

\bibitem [\protect \citeauthoryear {%
Hersbach%
\ \protect \BOthers {.}}{%
Hersbach%
\ \protect \BOthers {.}}{%
{\protect \APACyear {2020}}%
}]{%
hersbach2020era5}
\APACinsertmetastar {%
hersbach2020era5}%
\begin{APACrefauthors}%
Hersbach, H.%
, Bell, B.%
, Berrisford, P.%
, Hirahara, S.%
, Hor{\'a}nyi, A.%
, Mu{\~n}oz-Sabater, J.%
\BDBL {}others%
\end{APACrefauthors}%
\unskip\
\newblock
\APACrefYearMonthDay{2020}{}{}.
\newblock
{\BBOQ}\APACrefatitle {The ERA5 global reanalysis} {The era5 global
  reanalysis}.{\BBCQ}
\newblock
\APACjournalVolNumPages{Quarterly Journal of the Royal Meteorological
  Society}{146}{730}{1999--2049}.
\PrintBackRefs{\CurrentBib}

\bibitem [\protect \citeauthoryear {%
Hochreiter%
\ \BBA {} Schmidhuber%
}{%
Hochreiter%
\ \BBA {} Schmidhuber%
}{%
{\protect \APACyear {1997}}%
}]{%
hochreiter1997long}
\APACinsertmetastar {%
hochreiter1997long}%
\begin{APACrefauthors}%
Hochreiter, S.%
\BCBT {}\ \BBA {} Schmidhuber, J.%
\end{APACrefauthors}%
\unskip\
\newblock
\APACrefYearMonthDay{1997}{}{}.
\newblock
{\BBOQ}\APACrefatitle {Long short-term memory} {Long short-term memory}.{\BBCQ}
\newblock
\APACjournalVolNumPages{Neural computation}{9}{8}{1735--1780}.
\PrintBackRefs{\CurrentBib}

\bibitem [\protect \citeauthoryear {%
Hu%
, Chen%
, Wang%
\BCBL {}\ \BBA {} Li%
}{%
Hu%
\ \protect \BOthers {.}}{%
{\protect \APACyear {2022}}%
}]{%
hu2022swinvrnn}
\APACinsertmetastar {%
hu2022swinvrnn}%
\begin{APACrefauthors}%
Hu, Y.%
, Chen, L.%
, Wang, Z.%
\BCBL {}\ \BBA {} Li, H.%
\end{APACrefauthors}%
\unskip\
\newblock
\APACrefYearMonthDay{2022}{}{}.
\newblock
{\BBOQ}\APACrefatitle {SwinVRNN: A Data-Driven Ensemble Forecasting Model via
  Learned Distribution Perturbation} {Swinvrnn: A data-driven ensemble
  forecasting model via learned distribution perturbation}.{\BBCQ}
\newblock
\APACjournalVolNumPages{arXiv preprint arXiv:2205.13158}{}{}{}.
\PrintBackRefs{\CurrentBib}

\bibitem [\protect \citeauthoryear {%
Huang%
\ \protect \BOthers {.}}{%
Huang%
\ \protect \BOthers {.}}{%
{\protect \APACyear {2020}}%
}]{%
huang2020unet}
\APACinsertmetastar {%
huang2020unet}%
\begin{APACrefauthors}%
Huang, H.%
, Lin, L.%
, Tong, R.%
, Hu, H.%
, Zhang, Q.%
, Iwamoto, Y.%
\BDBL {}Wu, J.%
\end{APACrefauthors}%
\unskip\
\newblock
\APACrefYearMonthDay{2020}{}{}.
\newblock
{\BBOQ}\APACrefatitle {Unet 3+: A full-scale connected unet for medical image
  segmentation} {Unet 3+: A full-scale connected unet for medical image
  segmentation}.{\BBCQ}
\newblock
\BIn{} \APACrefbtitle {ICASSP 2020-2020 IEEE International Conference on
  Acoustics, Speech and Signal Processing (ICASSP)} {Icassp 2020-2020 ieee
  international conference on acoustics, speech and signal processing
  (icassp)}\ (\BPGS\ 1055--1059).
\PrintBackRefs{\CurrentBib}

\bibitem [\protect \citeauthoryear {%
Keisler%
}{%
Keisler%
}{%
{\protect \APACyear {2022}}%
}]{%
keisler2022forecasting}
\APACinsertmetastar {%
keisler2022forecasting}%
\begin{APACrefauthors}%
Keisler, R.%
\end{APACrefauthors}%
\unskip\
\newblock
\APACrefYearMonthDay{2022}{}{}.
\newblock
{\BBOQ}\APACrefatitle {Forecasting Global Weather with Graph Neural Networks}
  {Forecasting global weather with graph neural networks}.{\BBCQ}
\newblock
\APACjournalVolNumPages{arXiv preprint arXiv:2202.07575}{}{}{}.
\PrintBackRefs{\CurrentBib}

\bibitem [\protect \citeauthoryear {%
Kingma%
\ \BBA {} Ba%
}{%
Kingma%
\ \BBA {} Ba%
}{%
{\protect \APACyear {2014}}%
}]{%
kingma2014adam}
\APACinsertmetastar {%
kingma2014adam}%
\begin{APACrefauthors}%
Kingma, D\BPBI P.%
\BCBT {}\ \BBA {} Ba, J.%
\end{APACrefauthors}%
\unskip\
\newblock
\APACrefYearMonthDay{2014}{}{}.
\newblock
{\BBOQ}\APACrefatitle {Adam: A method for stochastic optimization} {Adam: A
  method for stochastic optimization}.{\BBCQ}
\newblock
\APACjournalVolNumPages{arXiv preprint arXiv:1412.6980}{}{}{}.
\PrintBackRefs{\CurrentBib}

\bibitem [\protect \citeauthoryear {%
Kipf%
\ \BBA {} Welling%
}{%
Kipf%
\ \BBA {} Welling%
}{%
{\protect \APACyear {2016}}%
}]{%
kipf2016semi}
\APACinsertmetastar {%
kipf2016semi}%
\begin{APACrefauthors}%
Kipf, T\BPBI N.%
\BCBT {}\ \BBA {} Welling, M.%
\end{APACrefauthors}%
\unskip\
\newblock
\APACrefYearMonthDay{2016}{}{}.
\newblock
{\BBOQ}\APACrefatitle {Semi-supervised classification with graph convolutional
  networks} {Semi-supervised classification with graph convolutional
  networks}.{\BBCQ}
\newblock
\APACjournalVolNumPages{arXiv preprint arXiv:1609.02907}{}{}{}.
\PrintBackRefs{\CurrentBib}

\bibitem [\protect \citeauthoryear {%
Krachmalnicoff%
\ \BBA {} Tomasi%
}{%
Krachmalnicoff%
\ \BBA {} Tomasi%
}{%
{\protect \APACyear {2019}}%
}]{%
krachmalnicoff2019convolutional}
\APACinsertmetastar {%
krachmalnicoff2019convolutional}%
\begin{APACrefauthors}%
Krachmalnicoff, N.%
\BCBT {}\ \BBA {} Tomasi, M.%
\end{APACrefauthors}%
\unskip\
\newblock
\APACrefYearMonthDay{2019}{}{}.
\newblock
{\BBOQ}\APACrefatitle {Convolutional neural networks on the HEALPix sphere: a
  pixel-based algorithm and its application to CMB data analysis}
  {Convolutional neural networks on the healpix sphere: a pixel-based algorithm
  and its application to cmb data analysis}.{\BBCQ}
\newblock
\APACjournalVolNumPages{Astronomy \& Astrophysics}{628}{}{A129}.
\PrintBackRefs{\CurrentBib}

\bibitem [\protect \citeauthoryear {%
Kurth%
\ \protect \BOthers {.}}{%
Kurth%
\ \protect \BOthers {.}}{%
{\protect \APACyear {2022}}%
}]{%
kurth2022fourcastnet}
\APACinsertmetastar {%
kurth2022fourcastnet}%
\begin{APACrefauthors}%
Kurth, T.%
, Subramanian, S.%
, Harrington, P.%
, Pathak, J.%
, Mardani, M.%
, Hall, D.%
\BDBL {}Anandkumar, A.%
\end{APACrefauthors}%
\unskip\
\newblock
\APACrefYearMonthDay{2022}{}{}.
\newblock
{\BBOQ}\APACrefatitle {Fourcastnet: Accelerating global high-resolution weather
  forecasting using adaptive fourier neural operators} {Fourcastnet:
  Accelerating global high-resolution weather forecasting using adaptive
  fourier neural operators}.{\BBCQ}
\newblock
\APACjournalVolNumPages{arXiv preprint arXiv:2208.05419}{}{}{}.
\PrintBackRefs{\CurrentBib}

\bibitem [\protect \citeauthoryear {%
Lam%
\ \protect \BOthers {.}}{%
Lam%
\ \protect \BOthers {.}}{%
{\protect \APACyear {2022}}%
}]{%
lam2022graphcast}
\APACinsertmetastar {%
lam2022graphcast}%
\begin{APACrefauthors}%
Lam, R.%
, Sanchez-Gonzalez, A.%
, Willson, M.%
, Wirnsberger, P.%
, Fortunato, M.%
, Pritzel, A.%
\BDBL {}others%
\end{APACrefauthors}%
\unskip\
\newblock
\APACrefYearMonthDay{2022}{}{}.
\newblock
{\BBOQ}\APACrefatitle {GraphCast: Learning skillful medium-range global weather
  forecasting} {Graphcast: Learning skillful medium-range global weather
  forecasting}.{\BBCQ}
\newblock
\APACjournalVolNumPages{arXiv preprint arXiv:2212.12794}{}{}{}.
\PrintBackRefs{\CurrentBib}

\bibitem [\protect \citeauthoryear {%
Li%
\ \protect \BOthers {.}}{%
Li%
\ \protect \BOthers {.}}{%
{\protect \APACyear {2020}}%
}]{%
li2020fourier}
\APACinsertmetastar {%
li2020fourier}%
\begin{APACrefauthors}%
Li, Z.%
, Kovachki, N.%
, Azizzadenesheli, K.%
, Liu, B.%
, Bhattacharya, K.%
, Stuart, A.%
\BCBL {}\ \BBA {} Anandkumar, A.%
\end{APACrefauthors}%
\unskip\
\newblock
\APACrefYearMonthDay{2020}{}{}.
\newblock
{\BBOQ}\APACrefatitle {Fourier neural operator for parametric partial
  differential equations} {Fourier neural operator for parametric partial
  differential equations}.{\BBCQ}
\newblock
\APACjournalVolNumPages{arXiv preprint arXiv:2010.08895}{}{}{}.
\PrintBackRefs{\CurrentBib}

\bibitem [\protect \citeauthoryear {%
Liu%
\ \protect \BOthers {.}}{%
Liu%
\ \protect \BOthers {.}}{%
{\protect \APACyear {2021}}%
}]{%
liu2021swin}
\APACinsertmetastar {%
liu2021swin}%
\begin{APACrefauthors}%
Liu, Z.%
, Lin, Y.%
, Cao, Y.%
, Hu, H.%
, Wei, Y.%
, Zhang, Z.%
\BDBL {}Guo, B.%
\end{APACrefauthors}%
\unskip\
\newblock
\APACrefYearMonthDay{2021}{}{}.
\newblock
{\BBOQ}\APACrefatitle {Swin transformer: Hierarchical vision transformer using
  shifted windows} {Swin transformer: Hierarchical vision transformer using
  shifted windows}.{\BBCQ}
\newblock
\BIn{} \APACrefbtitle {Proceedings of the IEEE/CVF International Conference on
  Computer Vision} {Proceedings of the ieee/cvf international conference on
  computer vision}\ (\BPGS\ 10012--10022).
\PrintBackRefs{\CurrentBib}

\bibitem [\protect \citeauthoryear {%
Liu%
\ \protect \BOthers {.}}{%
Liu%
\ \protect \BOthers {.}}{%
{\protect \APACyear {2022}}%
}]{%
liu2022convnet}
\APACinsertmetastar {%
liu2022convnet}%
\begin{APACrefauthors}%
Liu, Z.%
, Mao, H.%
, Wu, C\BHBI Y.%
, Feichtenhofer, C.%
, Darrell, T.%
\BCBL {}\ \BBA {} Xie, S.%
\end{APACrefauthors}%
\unskip\
\newblock
\APACrefYearMonthDay{2022}{}{}.
\newblock
{\BBOQ}\APACrefatitle {A convnet for the 2020s} {A convnet for the
  2020s}.{\BBCQ}
\newblock
\BIn{} \APACrefbtitle {Proceedings of the IEEE/CVF Conference on Computer
  Vision and Pattern Recognition} {Proceedings of the ieee/cvf conference on
  computer vision and pattern recognition}\ (\BPGS\ 11976--11986).
\PrintBackRefs{\CurrentBib}

\bibitem [\protect \citeauthoryear {%
Lopez-Gomez%
, McGovern%
, Agrawal%
\BCBL {}\ \BBA {} Hickey%
}{%
Lopez-Gomez%
\ \protect \BOthers {.}}{%
{\protect \APACyear {2022}}%
}]{%
lopez2022global}
\APACinsertmetastar {%
lopez2022global}%
\begin{APACrefauthors}%
Lopez-Gomez, I.%
, McGovern, A.%
, Agrawal, S.%
\BCBL {}\ \BBA {} Hickey, J.%
\end{APACrefauthors}%
\unskip\
\newblock
\APACrefYearMonthDay{2022}{}{}.
\newblock
{\BBOQ}\APACrefatitle {Global extreme heat forecasting using neural weather
  models} {Global extreme heat forecasting using neural weather models}.{\BBCQ}
\newblock
\APACjournalVolNumPages{arXiv preprint arXiv:2205.10972}{}{}{}.
\PrintBackRefs{\CurrentBib}

\bibitem [\protect \citeauthoryear {%
Lorenz%
}{%
Lorenz%
}{%
{\protect \APACyear {1969}}%
}]{%
lorenz69}
\APACinsertmetastar {%
lorenz69}%
\begin{APACrefauthors}%
Lorenz, E\BPBI N.%
\end{APACrefauthors}%
\unskip\
\newblock
\APACrefYearMonthDay{1969}{}{}.
\newblock
{\BBOQ}\APACrefatitle {The predictability of a flow which possesses many scales
  of motion} {The predictability of a flow which possesses many scales of
  motion}.{\BBCQ}
\newblock
\APACjournalVolNumPages{Tellus}{21}{3}{289--307}.
\PrintBackRefs{\CurrentBib}

\bibitem [\protect \citeauthoryear {%
Loshchilov%
\ \BBA {} Hutter%
}{%
Loshchilov%
\ \BBA {} Hutter%
}{%
{\protect \APACyear {2016}}%
}]{%
loshchilovsgdr}
\APACinsertmetastar {%
loshchilovsgdr}%
\begin{APACrefauthors}%
Loshchilov, I.%
\BCBT {}\ \BBA {} Hutter, F.%
\end{APACrefauthors}%
\unskip\
\newblock
\APACrefYearMonthDay{2016}{}{}.
\newblock
{\BBOQ}\APACrefatitle {SGDR: Stochastic Gradient Descent with Warm Restarts}
  {Sgdr: Stochastic gradient descent with warm restarts}.{\BBCQ}
\newblock
\BIn{} \APACrefbtitle {International Conference on Learning Representations.}
  {International conference on learning representations.}
\PrintBackRefs{\CurrentBib}

\bibitem [\protect \citeauthoryear {%
Palmer%
}{%
Palmer%
}{%
{\protect \APACyear {2019}}%
}]{%
palmer2019ecmwf}
\APACinsertmetastar {%
palmer2019ecmwf}%
\begin{APACrefauthors}%
Palmer, T.%
\end{APACrefauthors}%
\unskip\
\newblock
\APACrefYearMonthDay{2019}{}{}.
\newblock
{\BBOQ}\APACrefatitle {The ECMWF ensemble prediction system: Looking back (more
  than) 25 years and projecting forward 25 years} {The ecmwf ensemble
  prediction system: Looking back (more than) 25 years and projecting forward
  25 years}.{\BBCQ}
\newblock
\APACjournalVolNumPages{Quarterly Journal of the Royal Meteorological
  Society}{145}{}{12--24}.
\PrintBackRefs{\CurrentBib}

\bibitem [\protect \citeauthoryear {%
Pathak%
\ \protect \BOthers {.}}{%
Pathak%
\ \protect \BOthers {.}}{%
{\protect \APACyear {2022}}%
}]{%
pathak2022fourcastnet}
\APACinsertmetastar {%
pathak2022fourcastnet}%
\begin{APACrefauthors}%
Pathak, J.%
, Subramanian, S.%
, Harrington, P.%
, Raja, S.%
, Chattopadhyay, A.%
, Mardani, M.%
\BDBL {}others%
\end{APACrefauthors}%
\unskip\
\newblock
\APACrefYearMonthDay{2022}{}{}.
\newblock
{\BBOQ}\APACrefatitle {Fourcastnet: A global data-driven high-resolution
  weather model using adaptive fourier neural operators} {Fourcastnet: A global
  data-driven high-resolution weather model using adaptive fourier neural
  operators}.{\BBCQ}
\newblock
\APACjournalVolNumPages{arXiv preprint arXiv:2202.11214}{}{}{}.
\PrintBackRefs{\CurrentBib}

\bibitem [\protect \citeauthoryear {%
Perraudin%
, Defferrard%
, Kacprzak%
\BCBL {}\ \BBA {} Sgier%
}{%
Perraudin%
\ \protect \BOthers {.}}{%
{\protect \APACyear {2019}}%
}]{%
perraudin2019deepsphere}
\APACinsertmetastar {%
perraudin2019deepsphere}%
\begin{APACrefauthors}%
Perraudin, N.%
, Defferrard, M.%
, Kacprzak, T.%
\BCBL {}\ \BBA {} Sgier, R.%
\end{APACrefauthors}%
\unskip\
\newblock
\APACrefYearMonthDay{2019}{}{}.
\newblock
{\BBOQ}\APACrefatitle {DeepSphere: Efficient spherical convolutional neural
  network with HEALPix sampling for cosmological applications} {Deepsphere:
  Efficient spherical convolutional neural network with healpix sampling for
  cosmological applications}.{\BBCQ}
\newblock
\APACjournalVolNumPages{Astronomy and Computing}{27}{}{130--146}.
\PrintBackRefs{\CurrentBib}

\bibitem [\protect \citeauthoryear {%
Pfaff%
, Fortunato%
, Sanchez-Gonzalez%
\BCBL {}\ \BBA {} Battaglia%
}{%
Pfaff%
\ \protect \BOthers {.}}{%
{\protect \APACyear {2020}}%
}]{%
pfaff2020learning}
\APACinsertmetastar {%
pfaff2020learning}%
\begin{APACrefauthors}%
Pfaff, T.%
, Fortunato, M.%
, Sanchez-Gonzalez, A.%
\BCBL {}\ \BBA {} Battaglia, P\BPBI W.%
\end{APACrefauthors}%
\unskip\
\newblock
\APACrefYearMonthDay{2020}{}{}.
\newblock
{\BBOQ}\APACrefatitle {Learning mesh-based simulation with graph networks}
  {Learning mesh-based simulation with graph networks}.{\BBCQ}
\newblock
\APACjournalVolNumPages{arXiv preprint arXiv:2010.03409}{}{}{}.
\PrintBackRefs{\CurrentBib}

\bibitem [\protect \citeauthoryear {%
Rasp%
\ \protect \BOthers {.}}{%
Rasp%
\ \protect \BOthers {.}}{%
{\protect \APACyear {2023}}%
}]{%
rasp2023weatherbench}
\APACinsertmetastar {%
rasp2023weatherbench}%
\begin{APACrefauthors}%
Rasp, S.%
, Hoyer, S.%
, Merose, A.%
, Langmore, I.%
, Battaglia, P.%
, Russel, T.%
\BDBL {}others%
\end{APACrefauthors}%
\unskip\
\newblock
\APACrefYearMonthDay{2023}{}{}.
\newblock
{\BBOQ}\APACrefatitle {Weatherbench 2: A benchmark for the next generation of
  data-driven global weather models} {Weatherbench 2: A benchmark for the next
  generation of data-driven global weather models}.{\BBCQ}
\newblock
\APACjournalVolNumPages{arXiv preprint arXiv:2308.15560}{}{}{}.
\PrintBackRefs{\CurrentBib}

\bibitem [\protect \citeauthoryear {%
Rasp%
\ \BBA {} Thuerey%
}{%
Rasp%
\ \BBA {} Thuerey%
}{%
{\protect \APACyear {2021}}%
}]{%
rasp2021data}
\APACinsertmetastar {%
rasp2021data}%
\begin{APACrefauthors}%
Rasp, S.%
\BCBT {}\ \BBA {} Thuerey, N.%
\end{APACrefauthors}%
\unskip\
\newblock
\APACrefYearMonthDay{2021}{}{}.
\newblock
{\BBOQ}\APACrefatitle {Data-driven medium-range weather prediction with a
  resnet pretrained on climate simulations: A new model for weatherbench}
  {Data-driven medium-range weather prediction with a resnet pretrained on
  climate simulations: A new model for weatherbench}.{\BBCQ}
\newblock
\APACjournalVolNumPages{Journal of Advances in Modeling Earth
  Systems}{13}{2}{e2020MS002405}.
\PrintBackRefs{\CurrentBib}

\bibitem [\protect \citeauthoryear {%
Ronneberger%
, Fischer%
\BCBL {}\ \BBA {} Brox%
}{%
Ronneberger%
\ \protect \BOthers {.}}{%
{\protect \APACyear {2015}}%
}]{%
ronneberger2015u}
\APACinsertmetastar {%
ronneberger2015u}%
\begin{APACrefauthors}%
Ronneberger, O.%
, Fischer, P.%
\BCBL {}\ \BBA {} Brox, T.%
\end{APACrefauthors}%
\unskip\
\newblock
\APACrefYearMonthDay{2015}{}{}.
\newblock
{\BBOQ}\APACrefatitle {U-net: Convolutional networks for biomedical image
  segmentation} {U-net: Convolutional networks for biomedical image
  segmentation}.{\BBCQ}
\newblock
\BIn{} \APACrefbtitle {International Conference on Medical image computing and
  computer-assisted intervention} {International conference on medical image
  computing and computer-assisted intervention}\ (\BPGS\ 234--241).
\PrintBackRefs{\CurrentBib}

\bibitem [\protect \citeauthoryear {%
Scarselli%
, Gori%
, Tsoi%
, Hagenbuchner%
\BCBL {}\ \BBA {} Monfardini%
}{%
Scarselli%
\ \protect \BOthers {.}}{%
{\protect \APACyear {2008}}%
}]{%
scarselli2008graph}
\APACinsertmetastar {%
scarselli2008graph}%
\begin{APACrefauthors}%
Scarselli, F.%
, Gori, M.%
, Tsoi, A\BPBI C.%
, Hagenbuchner, M.%
\BCBL {}\ \BBA {} Monfardini, G.%
\end{APACrefauthors}%
\unskip\
\newblock
\APACrefYearMonthDay{2008}{}{}.
\newblock
{\BBOQ}\APACrefatitle {The graph neural network model} {The graph neural
  network model}.{\BBCQ}
\newblock
\APACjournalVolNumPages{IEEE transactions on neural networks}{20}{1}{61--80}.
\PrintBackRefs{\CurrentBib}

\bibitem [\protect \citeauthoryear {%
Scher%
\ \BBA {} Messori%
}{%
Scher%
\ \BBA {} Messori%
}{%
{\protect \APACyear {2018}}%
}]{%
scher2018predicting}
\APACinsertmetastar {%
scher2018predicting}%
\begin{APACrefauthors}%
Scher, S.%
\BCBT {}\ \BBA {} Messori, G.%
\end{APACrefauthors}%
\unskip\
\newblock
\APACrefYearMonthDay{2018}{}{}.
\newblock
{\BBOQ}\APACrefatitle {Predicting weather forecast uncertainty with machine
  learning} {Predicting weather forecast uncertainty with machine
  learning}.{\BBCQ}
\newblock
\APACjournalVolNumPages{Quarterly Journal of the Royal Meteorological
  Society}{144}{717}{2830--2841}.
\PrintBackRefs{\CurrentBib}

\bibitem [\protect \citeauthoryear {%
Scher%
\ \BBA {} Messori%
}{%
Scher%
\ \BBA {} Messori%
}{%
{\protect \APACyear {2019}}%
}]{%
Scher2019nn_GCM}
\APACinsertmetastar {%
Scher2019nn_GCM}%
\begin{APACrefauthors}%
Scher, S.%
\BCBT {}\ \BBA {} Messori, G.%
\end{APACrefauthors}%
\unskip\
\newblock
\APACrefYearMonthDay{2019}{}{}.
\newblock
{\BBOQ}\APACrefatitle {{Weather and climate forecasting with neural networks:
  using GCMs with different complexity as study-ground}} {{Weather and climate
  forecasting with neural networks: using GCMs with different complexity as
  study-ground}}.{\BBCQ}
\newblock
\APACjournalVolNumPages{Geoscientific Model Development}{12}{}{2797--2809}.
\PrintBackRefs{\CurrentBib}

\bibitem [\protect \citeauthoryear {%
Shen%
\ \protect \BOthers {.}}{%
Shen%
\ \protect \BOthers {.}}{%
{\protect \APACyear {2023}}%
}]{%
Shen:2023}
\APACinsertmetastar {%
Shen:2023}%
\begin{APACrefauthors}%
Shen, C.%
, Appling, A\BPBI P.%
, Gentine, P.%
, Bandai, T.%
, Gupta, H.%
, Tartakovsky, A.%
\BDBL {}Lawson, K.%
\end{APACrefauthors}%
\unskip\
\newblock
\APACrefYearMonthDay{2023}{}{}.
\newblock
{\BBOQ}\APACrefatitle {Differentiable modelling to unify machine learning and
  physical models for geosciences} {Differentiable modelling to unify machine
  learning and physical models for geosciences}.{\BBCQ}
\newblock
\APACjournalVolNumPages{Nature Reviews Earth \& Environment}{4}{8}{552--567}.
\newblock
\begin{APACrefURL} \url{https://doi.org/10.1038/s43017-023-00450-9}
  \end{APACrefURL}
\newblock
\begin{APACrefDOI} \doi{10.1038/s43017-023-00450-9} \end{APACrefDOI}
\PrintBackRefs{\CurrentBib}

\bibitem [\protect \citeauthoryear {%
Thuemmel%
\ \protect \BOthers {.}}{%
Thuemmel%
\ \protect \BOthers {.}}{%
{\protect \APACyear {2023}}%
}]{%
thuemmel2023inductive}
\APACinsertmetastar {%
thuemmel2023inductive}%
\begin{APACrefauthors}%
Thuemmel, J.%
, Karlbauer, M.%
, Otte, S.%
, Zarfl, C.%
, Martius, G.%
, Ludwig, N.%
\BDBL {}others%
\end{APACrefauthors}%
\unskip\
\newblock
\APACrefYearMonthDay{2023}{}{}.
\newblock
{\BBOQ}\APACrefatitle {Inductive biases in deep learning models for weather
  prediction} {Inductive biases in deep learning models for weather
  prediction}.{\BBCQ}
\newblock
\APACjournalVolNumPages{arXiv preprint arXiv:2304.04664}{}{}{}.
\PrintBackRefs{\CurrentBib}

\bibitem [\protect \citeauthoryear {%
Tobler%
}{%
Tobler%
}{%
{\protect \APACyear {1970}}%
}]{%
tobler1970computer}
\APACinsertmetastar {%
tobler1970computer}%
\begin{APACrefauthors}%
Tobler, W\BPBI R.%
\end{APACrefauthors}%
\unskip\
\newblock
\APACrefYearMonthDay{1970}{}{}.
\newblock
{\BBOQ}\APACrefatitle {A computer movie simulating urban growth in the Detroit
  region} {A computer movie simulating urban growth in the detroit
  region}.{\BBCQ}
\newblock
\APACjournalVolNumPages{Economic geography}{46}{sup1}{234--240}.
\PrintBackRefs{\CurrentBib}

\bibitem [\protect \citeauthoryear {%
Vaswani%
\ \protect \BOthers {.}}{%
Vaswani%
\ \protect \BOthers {.}}{%
{\protect \APACyear {2017}}%
}]{%
vaswani2017attention}
\APACinsertmetastar {%
vaswani2017attention}%
\begin{APACrefauthors}%
Vaswani, A.%
, Shazeer, N.%
, Parmar, N.%
, Uszkoreit, J.%
, Jones, L.%
, Gomez, A\BPBI N.%
\BDBL {}Polosukhin, I.%
\end{APACrefauthors}%
\unskip\
\newblock
\APACrefYearMonthDay{2017}{}{}.
\newblock
{\BBOQ}\APACrefatitle {Attention is all you need} {Attention is all you
  need}.{\BBCQ}
\newblock
\APACjournalVolNumPages{Advances in neural information processing
  systems}{30}{}{}.
\PrintBackRefs{\CurrentBib}

\bibitem [\protect \citeauthoryear {%
Vitart%
}{%
Vitart%
}{%
{\protect \APACyear {2004}}%
}]{%
Vitart2004}
\APACinsertmetastar {%
Vitart2004}%
\begin{APACrefauthors}%
Vitart, F.%
\end{APACrefauthors}%
\unskip\
\newblock
\APACrefYearMonthDay{2004}{}{}.
\newblock
{\BBOQ}\APACrefatitle {{Monthly forecasting at ECMWF}} {{Monthly forecasting at
  ECMWF}}.{\BBCQ}
\newblock
\APACjournalVolNumPages{Monthly Weather Review}{132}{}{2761--2779}.
\newblock
\begin{APACrefDOI} \doi{10.1175/MWR2826.1} \end{APACrefDOI}
\PrintBackRefs{\CurrentBib}

\bibitem [\protect \citeauthoryear {%
Weigel%
, Baggenstos%
, Liniger%
, Vitart%
\BCBL {}\ \BBA {} Appenzeller%
}{%
Weigel%
\ \protect \BOthers {.}}{%
{\protect \APACyear {2008}}%
}]{%
Weigel2008}
\APACinsertmetastar {%
Weigel2008}%
\begin{APACrefauthors}%
Weigel, A\BPBI P.%
, Baggenstos, D.%
, Liniger, M\BPBI A.%
, Vitart, F.%
\BCBL {}\ \BBA {} Appenzeller, C.%
\end{APACrefauthors}%
\unskip\
\newblock
\APACrefYearMonthDay{2008}{}{}.
\newblock
{\BBOQ}\APACrefatitle {{Probabilistic Verification of Monthly Temperature
  Forecasts}} {{Probabilistic Verification of Monthly Temperature
  Forecasts}}.{\BBCQ}
\newblock
\APACjournalVolNumPages{Monthly Weather Review}{136}{}{5162--5182}.
\newblock
\begin{APACrefDOI} \doi{10.1175/2008MWR2551.1} \end{APACrefDOI}
\PrintBackRefs{\CurrentBib}

\bibitem [\protect \citeauthoryear {%
Weyn%
, Durran%
\BCBL {}\ \BBA {} Caruana%
}{%
Weyn%
\ \protect \BOthers {.}}{%
{\protect \APACyear {2019}}%
}]{%
weyn2019can}
\APACinsertmetastar {%
weyn2019can}%
\begin{APACrefauthors}%
Weyn, J\BPBI A.%
, Durran, D\BPBI R.%
\BCBL {}\ \BBA {} Caruana, R.%
\end{APACrefauthors}%
\unskip\
\newblock
\APACrefYearMonthDay{2019}{}{}.
\newblock
{\BBOQ}\APACrefatitle {Can machines learn to predict weather? Using deep
  learning to predict gridded 500-hPa geopotential height from historical
  weather data} {Can machines learn to predict weather? using deep learning to
  predict gridded 500-hpa geopotential height from historical weather
  data}.{\BBCQ}
\newblock
\APACjournalVolNumPages{Journal of Advances in Modeling Earth
  Systems}{11}{8}{2680--2693}.
\PrintBackRefs{\CurrentBib}

\bibitem [\protect \citeauthoryear {%
Weyn%
, Durran%
\BCBL {}\ \BBA {} Caruana%
}{%
Weyn%
\ \protect \BOthers {.}}{%
{\protect \APACyear {2020}}%
}]{%
weyn2020improving}
\APACinsertmetastar {%
weyn2020improving}%
\begin{APACrefauthors}%
Weyn, J\BPBI A.%
, Durran, D\BPBI R.%
\BCBL {}\ \BBA {} Caruana, R.%
\end{APACrefauthors}%
\unskip\
\newblock
\APACrefYearMonthDay{2020}{}{}.
\newblock
{\BBOQ}\APACrefatitle {Improving data-driven global weather prediction using
  deep convolutional neural networks on a cubed sphere} {Improving data-driven
  global weather prediction using deep convolutional neural networks on a cubed
  sphere}.{\BBCQ}
\newblock
\APACjournalVolNumPages{Journal of Advances in Modeling Earth
  Systems}{12}{9}{e2020MS002109}.
\PrintBackRefs{\CurrentBib}

\bibitem [\protect \citeauthoryear {%
Weyn%
, Durran%
, Caruana%
\BCBL {}\ \BBA {} Cresswell-Clay%
}{%
Weyn%
\ \protect \BOthers {.}}{%
{\protect \APACyear {2021}}%
}]{%
weyn2021sub}
\APACinsertmetastar {%
weyn2021sub}%
\begin{APACrefauthors}%
Weyn, J\BPBI A.%
, Durran, D\BPBI R.%
, Caruana, R.%
\BCBL {}\ \BBA {} Cresswell-Clay, N.%
\end{APACrefauthors}%
\unskip\
\newblock
\APACrefYearMonthDay{2021}{}{}.
\newblock
{\BBOQ}\APACrefatitle {Sub-seasonal forecasting with a large ensemble of
  deep-learning weather prediction models} {Sub-seasonal forecasting with a
  large ensemble of deep-learning weather prediction models}.{\BBCQ}
\newblock
\APACjournalVolNumPages{Journal of Advances in Modeling Earth
  Systems}{13}{7}{e2021MS002502}.
\PrintBackRefs{\CurrentBib}

\bibitem [\protect \citeauthoryear {%
Zhou%
, Rahman~Siddiquee%
, Tajbakhsh%
\BCBL {}\ \BBA {} Liang%
}{%
Zhou%
\ \protect \BOthers {.}}{%
{\protect \APACyear {2018}}%
}]{%
zhou2018unet++}
\APACinsertmetastar {%
zhou2018unet++}%
\begin{APACrefauthors}%
Zhou, Z.%
, Rahman~Siddiquee, M\BPBI M.%
, Tajbakhsh, N.%
\BCBL {}\ \BBA {} Liang, J.%
\end{APACrefauthors}%
\unskip\
\newblock
\APACrefYearMonthDay{2018}{}{}.
\newblock
{\BBOQ}\APACrefatitle {Unet++: A nested u-net architecture for medical image
  segmentation} {Unet++: A nested u-net architecture for medical image
  segmentation}.{\BBCQ}
\newblock
\BIn{} \APACrefbtitle {Deep learning in medical image analysis and multimodal
  learning for clinical decision support} {Deep learning in medical image
  analysis and multimodal learning for clinical decision support}\ (\BPGS\
  3--11).
\newblock
\APACaddressPublisher{}{Springer}.
\PrintBackRefs{\CurrentBib}

\end{thebibliography}

%
%
%
%
%

\end{document}